\documentclass[a4paper,11pt]{article}
\usepackage{jheppub} 
\usepackage[T1]{fontenc} 
\usepackage{epsfig}

\preprint{IPHC-PHENO-12-07, MS-TP-12-05}

\title{Gaugino production in proton-proton collisions \\ at a center-of-mass energy of 8 TeV}
\author[a]{Benjamin Fuks,}
\author[b]{Michael Klasen,}
\author[b]{David R.\ Lamprea}
\author[b]{and Marcel Rothering}

\affiliation[a]{Institut Pluridisciplinaire Hubert Curien/D\'epartement
  Recherches Subatomiques, Universit\'e de Strasbourg/CNRS-IN2P3,
  23 rue du Loess, F-67037 Strasbourg, France}
\affiliation[b]{Institut f\"ur Theoretische Physik, Westf\"alische
  Wilhelms-Universit\"at M\"unster,\\ 
  Wilhelm-Klemm-Stra\ss{}e 9, D-48149 M\"unster, Germany}

\emailAdd{benjamin.fuks@iphc.cnrs.fr}
\emailAdd{michael.klasen@uni-muenster.de}
\emailAdd{david.lamprea@uni-muenster.de}
\emailAdd{marcel.rothering@uni-muenster.de}

\abstract{
 Motivated by hints for a light Standard Model-like
 Higgs boson and a shift in experimental attention towards
 electroweak supersymmetry particle production at the CERN LHC, we update in this paper
 our precision predictions at next-to-leading order of perturbative QCD matched to
 resummation at the next-to-leading logarithmic accuracy for direct gaugino pair production
 in proton-proton collisions with a center-of-mass energy of 8 TeV. Tables of total
 cross sections are presented together with the corresponding scale and parton density
 uncertainties for benchmark points adopted recently by the experimental collaborations,
 and figures are presented for up-to-date model lines attached to them. 
 Since the experimental analyses are currently obtained with parton showers matched
 to multi-parton matrix elements, we also analyze the precision of this procedure by
 comparing invariant-mass and transverse-momentum distributions obtained in this way
 to those obtained with threshold and transverse-momentum resummation.
}

\keywords{Resummation, supersymmetry, hadron collider phenomenology,
superparticle searches}

\begin{document}

\def\beq{\begin{equation}}
\def\eeq{\end{equation}}
\def\bea{\begin{eqnarray}}
\def\eea{\end{eqnarray}}
\def\bsp#1\esp{\begin{split}#1\end{split}}

\def\d{{\rm d}}
\def\eps{\epsilon}
\def\mgl{m_{\tilde g}}
\def\msq{m_{\tilde q}}
\newcommand{\neu}[1]{\tilde \chi^0_{#1}}
\def\tb{\tan\beta}

\def\la{\langle }
\def\ra{\rangle }
\def\lp{\left. }
\def\rp{\right. }
\def\lr{\left( }
\def\rr{\right) }
\def\le{\left[ }
\def\re{\right] }
\def\lg{\left\{ }
\def\rg{\right\} }
\def\lb{\left| }
\def\rb{\right| }

\maketitle
\flushbottom


\section{Introduction}
\label{sec:1}

For many years, the Standard Model (SM) of particle physics has passed all experimental
tests. Only the mechanism of electroweak symmetry breaking and of the generation
of mass has remained as an unsolved question and is now being addressed with the
general-purpose experiments ATLAS and CMS at the CERN LHC. The discovery of a Standard
Model-like scalar Higgs boson of mass of about 125 GeV and first measurements
of its decay channels now seem imminent \cite{ATLAS:2012ae,Chatrchyan:2012tx},
which would represent an impressive success of this large technical endeavor. However,
the mass of a fundamental scalar particle is affected by large quantum corrections,
so that immediately the question of its stabilization with respect to the Planck
scale arises.

Weak-scale supersymmetry (SUSY) \cite{Nilles:1983ge,Haber:1984rc} has long been
known to effectively solve not only this ``hierarchy problem'', but also a considerable
number of other problems such as the Grand Unification of gauge symmetries and the
question of a viable dark matter candidate. LHC searches for supersymmetric particles
have concentrated up to now on the largest, strong production channels, but so far no
signs of squarks and gluinos have been detected \cite{Lowette:2012uh}. Based on LHC
searches with about 1 fb$^{-1}$ of data, benchmark points have therefore been defined
for future SUSY searches at the LHC \cite{AbdusSalam:2011fc}, and the strong production
cross sections and their theoretical uncertainties have been updated to next-to-leading
order (NLO) and next-to-leading logarithmic (NLL) accuracy for $pp$ collisions with
a center-of-mass energy of 7 TeV \cite{Kulesza:2008jb,Kulesza:2009kq,
Beenakker:2009ha, Beenakker:2010nq, Beenakker:2011fu, Kramer:2012bx}.

In 2012, the LHC center-of-mass energy has increased to 8 TeV. In addition,
the negative search results for gluino and degenerate squark production and their
cascade decays \cite{Lowette:2012uh} imply that the experimental attention now shifts
towards third-generation squarks and the weak production channels, in particular the
direct production of gauginos. A first investigation of neutralino-chargino pair production
and their ``golden'' decay into three charged leptons has already been published
by the ATLAS collaboration with the result that in simplified models such as the
constrained Minimal Supersymmetric Standard Model (cMSSM), degenerate lightest
chargino
($\tilde{\chi}^\pm_1$) and next-to-lightest neutralino ($\tilde{\chi}^0_2$) masses are
excluded up to 300 GeV, depending on the mass of the lightest SUSY particle (LSP)
assumed to be the lightest neutralino $\tilde{\chi}^0_1$ \cite{:2012cwa}. Surprisingly,
an independent analysis of ATLAS and CMS measurements of the $W^+W^-$ cross section seems
to fit theoretical predictions better when electroweak gauginos of about 100 GeV
are included than by the SM cross section alone \cite{Curtin:2012nn}.

The observations above motivate us to update in this paper our precision (NLO+NLL)
predictions for direct gaugino pair production. They have been obtained by combining
a leading order (LO) calculation \cite{Debove:2008nr} and its next-to-leading order corrections
\cite{Debove:2010kf}, which had already been computed independently previously
\cite{Beenakker:1999xh}, with resummation of the leading and next-to-leading logarithms
to all orders in the threshold \cite{Debove:2010kf} and small transverse-momentum ($p_T$)
\cite{Debove:2009ia} regimes or simultaneously in both \cite{Debove:2011xj}.
A similar program has been carried out for the other weak channel, {\em i.e.} the
production of sleptons \cite{Bozzi:2004qq,Bozzi:2006fw,Bozzi:2007qr,Bozzi:2007tea}.
However, their production cross sections are considerably smaller and are not yet
accessible at the LHC, so that we leave the corresponding theoretical update for
future work. The semiweak associated production of gauginos with squarks
\cite{Plehn:2004rp,Binoth:2011xi} and gluinos \cite{Berger:1999mc,Berger:2000iu,Spira:2002rd},
which may well become important should the energy available at the LHC prove insufficient
for the pair production of the latter, has so far been computed only up to NLO.

In Section \ref{sec:2} we present, for a representative selection of benchmark points
defined in Ref.\ \cite{AbdusSalam:2011fc}, the total gaugino pair production cross
sections and the corresponding scale and parton density function (PDF) uncertainties
at the current center-of-mass energy of 8 TeV in tabular form. For comparison
and analyses of the data taken previously at 7 TeV, we list the corresponding cross
sections and uncertainties in Appendix \ref{sec:a}. In addition, an overview of the
absolute size of the total cross sections at 8 TeV, the impact of the higher-order
corrections, and the theoretical uncertainties can be obtained from the figures
shown in Section \ref{sec:2} illustrating the cross sections as a function of
the physical gaugino mass along the model lines attached to the selected
benchmark points.

Since the experimental analyses \cite{:2012cwa} are currently based on LO Monte
Carlo simulations of the gaugino signal with parton showers \cite{Corcella:2000bw,%
Sjostrand:2006za}, matched \cite{Mangano:2002ea} to hard tree-level multi-parton
corrections \cite{Alwall:2011uj} and normalized to the NLO total cross section
\cite{Beenakker:1999xh}, we take the opportunity to analyze in this paper the
precision of this procedure. We therefore compare in Section \ref{sec:3}
invariant-mass and transverse-momentum distributions obtained with parton showers
only \cite{Sjostrand:2006za} and matched \cite{Mangano:2002ea} to one or two
additional hard jets \cite{Alwall:2011uj} to those obtained with 
threshold \cite{Debove:2010kf} and transverse-momentum resummation
\cite{Debove:2009ia}. Similar comparisons have been presented previously for the
hadroproduction of new neutral gauge ($Z'$) bosons and charged Higgs ($H^\pm$) bosons
in association with top quarks in Refs.\ \cite{Fuks:2007gk} and
\cite{Weydert:2009vr,Klasen:2012wq}, respectively.

Finally, we summarize our results and present our conclusions in Section
\ref{sec:4}.

\section{Total cross sections at $\sqrt{s}=8$ TeV}
\label{sec:2}

As the ATLAS and CMS experiments are now starting to probe the direct production
of gaugino pairs decaying, {\em e.g.}, into three charged leptons \cite{:2012cwa},
we present in this Section an illustrative selection of total gaugino cross sections at
the current center-of-mass energy of $\sqrt{s}=8$ TeV in order to facilitate precise
comparisons with the experimental results. The corresponding cross sections
at $\sqrt{s}=7$ TeV, relevant for data taken in 2010 and 2011, can be found in
Appendix \ref{sec:a}. Further results are available from the authors upon request.

\subsection{Benchmark points}
\label{sec:2a}

First SUSY search results at the LHC with an integrated luminosity of about 
1 fb$^{-1}$ of data excluded six out of the nine SPS benchmark points that had
been in use since about a decade \cite{Allanach:2002nj,AguilarSaavedra:2005pw}.
These cMSSM points therefore had to be replaced by
new lines \cite{AbdusSalam:2011fc} where, motivated by the
constraints derived from the measurements of the anomalous magnetic moment of
the muon $(g-2)_\mu$ and the rare $b\to s\gamma$ decay, only models with a
positive off-diagonal Higgs mixing parameter $\mu>0$ and $\tan\beta=10$, $A_0=0$
GeV or $\tan\beta=40$, $A_0=-500$ GeV were chosen. Here, $\tan\beta$
denotes the ratio of the vacuum
expectation values (VEVs) of the neutral components of the two Higgs doublets, and 
$A_0$ is the universal soft trilinear coupling of the Higgs fields to the
squarks at the Grand Unification scale.
In discussions among the SUSY working groups of the ATLAS and CMS
experiments and with the LHC Physics Center at CERN (LPCC) \cite{lpcc}, 49
benchmark points lying on these lines were retained for detailed cross section
studies \cite{lhcsusy}.

To keep the number of figures manageable, we have selected thirteen out of these
49 points which are included in the two model lines 10.1 and 10.3 with
$\tan\beta=10$ and $A_0=0$ GeV as well as seven additional points located on
a line similar to line 40.1 with
$\tan\beta=40$ and $A_0=-500$ GeV. Concerning the tables with detailed cross
section analyses presented in Section \ref{sec:2b} and the comparison with the
Monte Carlo
predictions investigated in Section \ref{sec:3}, we decide to focus in contrast
on one 
specific benchmark scenario for each of the selected lines. The latter are
listed in bold face in Tables \ref{tab:1}, \ref{tab:2} and \ref{tab:3}, where we
employ the LPCC numbering scheme.

%
\begin{table}[t]
\begin{center}
\begin{tabular}{|c|c||c|c||c|}
\hline
 Point & $(m_{1/2}, m_0)$ [GeV]& $\mgl$ [GeV] & $\langle \msq \rangle$
 [GeV] & BR$(\neu{2} \to {\tilde \ell}\ell / {\tilde \tau}\tau)$ [\%]\\
\hline
{\bf 1} & {\bf (400, 100)}   & {\bf 935} & {\bf 840} & {\bf 16} / {\bf 22} \\
2 & (450, 112.5) & 1040 & 940 & 20 / 20 \\
3 & (500, 125) & 1145 & 1030 & 24 / 19 \\
4 & (550, 137.5) & 1255 & 1125 & 26 / 18 \\
5 & (600, 150) & 1355 & 1220 & 28 / 18 \\
6 & (650, 162.5) & 1460 & 1310 & 28 / 17 \\
7 & (700, 175) & 1565 & 1405 & 29 / 17 \\
\hline
\end{tabular}
\caption{\label{tab:1}
 Selection of benchmark points on the cMSSM model line 10.1 of Ref.\
 \cite{AbdusSalam:2011fc} with $\tb = 10$, $A_0 = 0$ GeV, and $m_0 = 0.25 \times
 m_{1/2}$. The points are spaced in steps of
 $\Delta m_{1/2} = 50$ GeV, and the gluino and average squark masses are rounded
 to 5 GeV accuracy. We also present the branching ratios of the
 next-to-lightest neutralino into $\ell=e,\mu$ and $\tau$ (s)leptons.}
\end{center}
\end{table}
%

For the first point, with a low value of $\tan\beta=10$ and a vanishing
universal trilinear coupling $A_0 = 0$ GeV, we have chosen an optimistic
scenario with low 
values for the universal scalar mass $m_0$ and for the universal gaugino mass
$m_{1/2}$ at the high scale, the point 1 of the LPCC numbering scheme. These
values of $m_0$ and $m_{1/2}$ lead to modest neutralino and chargino 
masses at the electroweak scale of about 150--550 GeV (see Table \ref{tab:4}
in Section \ref{sec:2b}). In Table \ref{tab:1} we also list,
together with the
universal scalar and gaugino masses, the physical gluino and average squark
masses, which lie in the 1--2 TeV range and are 
therefore compliant with the current experimental bounds from the ATLAS
and CMS experiments. The associated production of the lightest chargino
$\tilde\chi_1^\pm$ with the next-to-lightest neutralino $\tilde\chi_2^0$
leads often to the golden trilepton signature. We therefore also present in Table
\ref{tab:1} the branching ratios of the second lightest neutralino to first and
second generation and (s)tau (s)leptons.
These quantities indeed allow us to deduce the trilepton production rate from
the total cross sections given in the next subsections, since for all the
scenarios lying on the line 10.1 the
lightest chargino $\tilde\chi_1^\pm$ and the sleptons always decay (in a first
approximation) to a single observable lepton and missing transverse energy.

%
\begin{table}[t]
\begin{center}
\begin{tabular}{|c||c||c|c||c|}
\hline
Point & $(m_{1/2}, m_0)$ [GeV] & $\mgl$ [GeV] & $\langle \msq \rangle$
[GeV] & BR$(\neu{2} \to \neu{1} h)$ [\%] \\
\hline
15 & (375, 250) & 885 & 825 & 89 \\
16 & (450, 300) & 1050 & 975 & 92 \\
17 & (525, 350) & 1210 & 1125 & 92 \\
{\bf 18} & {\bf (600, 400)} & {\bf 1370} & {\bf 1275} & {\bf 92} \\
19 & (675, 450) & 1525 & 1420 &  92 \\
20 & (750, 500) & 1680 & 1565 &  92 \\
\hline
\end{tabular}
\caption{\label{tab:2}
 Selection of benchmark points on the cMSSM model line 10.3 of Ref.\
 \cite{AbdusSalam:2011fc} with $\tb = 10$, $A_0 = 0$ GeV, and $m_{1/2} = 1.5 \times
 m_0$. The points are spaced in steps of
 $\Delta m_0 = 50$ GeV, and the gluino and average squark masses are rounded
 to 5 GeV accuracy. We also present the branching ratio of the dominant decay mode of the
 next-to-lightest neutralino into a Higgs boson.}
\end{center}
\end{table}
%

Our second benchmark scenario consists in the point 18 lying on the model line
10.3. As for the point 1, the value of the ratio of the Higgs VEVs has been set
to $\tan\beta=10$ and the universal trilinear scalar coupling is vanishing at
high energies. In contrast to the first scenario, we rather adopt here higher 
values for the universal scalar and
gaugino masses $m_0$ and $m_{1/2}$, which yields slightly heavier neutralino and
chargino states with masses of about 250--770 GeV (see Table \ref{tab:5} in
Section \ref{sec:2b}). As for
the first scenario, this combination of values for $m_0$ and $m_{1/2}$ shifts
the masses of the squarks and gluino to the 1--2 TeV range as shown
in Table \ref{tab:2}, rendering the scenarios of the line 10.3 not (yet)
excluded by LHC data. As shown in the last column of Table \ref{tab:2}, the
next-to-lightest neutralino decays most of the time to a Higgs boson, 
since as an almost pure light wino it can not decay into the
heavier superpartners of the left-handed leptons and quarks.
As another consequence of the heavy squark masses, $t$- and $u$-channel squark
exchanges leading to the production of neutralino and gaugino pairs are
suppressed by heavy propagators. One however expects reasonable
production rates for final states containing one or two neutralinos
$\tilde\chi_2^0$ via dominant
$s$-channel weak boson exchanges and reduced destructive interferences with the
$t$- and $u$-channel diagrams (in contrast to a light squark scenario). Since
the
lightest chargino decays with a 98\% branching ratio to a $W$-boson, one can
deduce the production rate of golden signatures from the values of the
$\tilde\chi_2^0\tilde\chi_1^\pm$ total cross section presented in Section
\ref{sec:2b} and Section \ref{sec:2c} after accounting for the branching ratio
of a Higgs boson decaying
to a pair of tau leptons and the rate of leptonic $W$-boson decays.

%
\begin{table}[t]
\begin{center}

\renewcommand{\arraystretch}{1.2}
\begin{tabular}{|c||c||c|c||c|}
\hline
 Point & $(m_{1/2}, m_0)$ [GeV]& $\mgl$ [GeV]& $\langle \msq \rangle$ / 
 $m_{\tilde t_1}$ / $m_{\tilde b_1}$ [GeV] & BR$(\neu{2} \!\to\!
 {\tilde \tau_1}\tau/ h\tilde\chi_1^0) [\%]$\\
\hline
27 & (400, 300) & 940  & 890  / 615 / 745 & 98/2\\
28 & (450, 325) & 1050 & 985  / 700 / 835 & 97/3\\
29 & (500, 350) & 1155 & 1185 / 780 / 925 & 96/4\\
30 & (550, 375) & 1260 & 1180 / 860 / 1010 & 95/4\\
{\bf 31} & {\bf (600, 400)} & {\bf 1365} & {\bf 1275} / {\bf 940} / {\bf
1100} & {\bf 95/5}\\
32 & (650, 425) & 1470 & 1370 / 1020 / 1185 & 94/5 \\
33 & (700, 450) & 1575 & 1465 / 1095 / 1275 & 94/6 \\
\hline
\end{tabular}
\renewcommand{\arraystretch}{1}
\caption{\label{tab:3}
 Selection of benchmark points on a cMSSM model line similar to line 40.1 of Ref.\
 \cite{AbdusSalam:2011fc} with $\tb = 40$, $A_0 = -500$ GeV, and $m_0 = 0.5 \times
 m_{1/2} + 100$ GeV. The points are spaced in steps of
 $\Delta m_{1/2} = 50$ GeV. The mass of the gluino, the average mass of
 the heavier squarks, and the masses of the (considerably lighter) top and
 bottom squarks are rounded to 5 GeV accuracy. We also present the branching
 ratios for the decays of the next-to-lightest neutralino into $\tau$ (s)leptons
 and Higgs bosons.}
\end{center}
\end{table}
%

Our third selected scenario lies on a model line similar to the line 40.1 of Ref.\
\cite{AbdusSalam:2011fc} with a large value of $\tan\beta=40$ and a large
negative value of the universal scalar trilinear coupling $A_0=-500$ GeV. We
retain the
high values of the universal scalar and gaugino masses that have driven the
choice of our second scenario and therefore adopt the point 31 of the LPCC
numbering scheme as our third benchmark point. Comparing with the point 18, the
neutralino and chargino masses are not drastically affected by the different
choices of $\tan\beta$ and $A_0$ and hence lie in the 250--825 GeV range (see
Table \ref{tab:6} in Section \ref{sec:2b}). Similarly, the gluino and first and
second generation squark
masses are not drastically affected either as presented in Table \ref{tab:3}. In
contrast, the large values of $\tan\beta$ and $A_0$ induce an important mixing
among the third-generation squark interaction eigenstates (and to a smaller extent
among the left- and right-handed stau states). As a consequence, the masses of
the lightest stop and sbottom are much lower than the average squark mass as
shown in Table \ref{tab:3}, and these states are almost maximal
admixtures of the left-handed and right-handed third generation squark
eigenstates. We also present the
two main decay modes of the second neutralino, decaying mainly to an associated
pair of stau (always decaying to a tau and a lightest neutralino) and 
tau lepton but also, at a much smaller rate, to a Higgs boson and missing energy
carried by the lightest neutralino.
As for the two other selected scenarios, the cross section
corresponding to the trilepton mode can be deduced from the values of the total
cross sections given in Section \ref{sec:2b} and Section \ref{sec:2c} 
and from the branching ratios of the lightest Higgs boson to taus as well as the
one associated to tau leptonic decays.

We emphasize that while the increasing experimental limits on the squark and
gluino masses might exclude 
the corresponding cMSSM scenarios, the (lower) gaugino masses may still
remain allowed and their total cross sections therefore approximately valid
in the context of more general SUSY-breaking scenarios.

\subsection{Total cross sections for the benchmark points 1, 18, and 31}
\label{sec:2b}

The production of pairs of charginos and neutralinos has been initially studied
at leading order of perturbative QCD in the early 1980s \cite{Barger:1983wc,
Dawson:1983fw}, while more recently polarization  \cite{Debove:2008nr,
Klasen:2010bk} and flavor-violating \cite{Bozzi:2007me, delAguila:2008iz,
Fuks:2008ab} effects have been included and investigated. Next-to-leading
order corrections have been supplemented to the leading order approximation 
\cite{Beenakker:1999xh, Debove:2010kf} and have been found to be
important, in particular due to the presence of large logarithmic
contributions arising from soft and collinear parton emission by the initial
state particles. Since they spoil the convergence of the perturbative series, these
logarithms have to be resummed to all orders in the strong coupling constant
to allow for reliable theoretical predictions in the entire phase space,
including the regions related to soft and collinear QCD radiation.
Transverse-momentum and threshold resummation have then been achieved at the
next-to-leading logarithmic accuracy independently \cite{Debove:2010kf,
Debove:2009ia} and simultaneously \cite{Debove:2011xj}.

In this Section, we present in Tables \ref{tab:4}, \ref{tab:5}, and \ref{tab:6}
total production cross sections for various combinations of neutralino and
chargino pairs in the context of the benchmark points 1, 18 and 31 introduced in
Section \ref{sec:2a} at the leading and next-to-leading order of perturbative
QCD as well as after matching the NLO results to threshold resummation. The
masses of the produced gauginos are also given as references, while the values
of the squark masses entering the partonic cross section can be found in Tables
\ref{tab:1}, \ref{tab:2} and \ref{tab:3}.

%
\begin{table}[!t]
\renewcommand{\arraystretch}{1.2}
\begin{center}
\begin{tabular}{| c || l | l || l | l | l |}
\hline
 Process & $m_1$ [GeV] & $m_2$ [GeV] &  LO [fb] & NLO [fb] & NLO+NLL [fb] \\
\hline
$p p \to \chi^0_1 \chi^0_1$ & 161.7 & 161.7 & $0.81^{+5.8 \%}_{-5.3 \%}$  &  $1.06^{+3.5 \%}_{-3.0 \%}{}^{+2.8 \%}_{-2.0 \%}$  & $1.03^{+0.5 \%}_{-0.6 \%}{}^{+2.9 \%}_{-2.0 \%}$\\
$p p \to \chi^0_1 \chi^-_1$ & 161.7 & 303.5 & $0.16^{+6.0 \%}_{-5.5 \%}$  &  $0.20^{+2.5 \%}_{-2.4 \%}{}^{+2.9 \%}_{-2.4 \%}$  & $0.20^{+0.0 \%}_{-0.3 \%}{}^{+2.9 \%}_{-2.5 \%}$\\
$p p \to \chi^0_2 \chi^0_2$ & 303.8 & 303.8 & $0.85^{+9.2 \%}_{-7.9 \%}$  &  $1.07^{+3.5 \%}_{-3.5 \%}{}^{+3.1 \%}_{-2.2 \%}$  & $1.05^{+0.0 \%}_{-0.4 \%}{}^{+3.5 \%}_{-1.9 \%}$\\
$p p \to \chi^0_2 \chi^0_3$ & 303.8 & 526.5 & $0.21^{+9.4 \%}_{-8.1 \%}$  &  $0.25^{+2.6 \%}_{-2.9 \%}{}^{+3.2 \%}_{-2.3 \%}$  & $0.25^{+0.1 \%}_{-0.5 \%}{}^{+3.2 \%}_{-2.3 \%}$\\
$p p \to \chi^0_2 \chi^-_1$ & 303.8 & 303.5 & $14.46^{+6.7 \%}_{-6.1 \%}$  &  $17.25^{+1.6 \%}_{-1.7 \%}{}^{+3.0 \%}_{-2.6 \%}$  & $17.05^{+0.2 \%}_{-0.7 \%}{}^{+3.1 \%}_{-2.6 \%}$\\
$p p \to \chi^0_3 \chi^0_4$ & 526.5 & 542.4 & $0.83^{+11.0 \%}_{-9.3 \%}$  &  $0.97^{+2.8 \%}_{-3.3 \%}{}^{+3.9 \%}_{-2.4 \%}$  & $0.96^{+0.4 \%}_{-0.9 \%}{}^{+3.8 \%}_{-2.5 \%}$\\
$p p \to \chi^0_3 \chi^-_1$ & 526.5 & 303.5 & $0.12^{+9.4 \%}_{-8.1 \%}$  &  $0.15^{+2.6 \%}_{-2.9 \%}{}^{+3.8 \%}_{-2.9 \%}$  & $0.15^{+0.1 \%}_{-0.6 \%}{}^{+3.8 \%}_{-3.0 \%}$\\
$p p \to \chi^0_3 \chi^-_2$ & 526.5 & 542.2 & $0.42^{+11.2 \%}_{-9.5 \%}$  &  $0.50^{+2.8 \%}_{-3.3 \%}{}^{+4.9 \%}_{-3.6 \%}$  & $0.49^{+0.4 \%}_{-0.9 \%}{}^{+4.9 \%}_{-3.5 \%}$\\
$p p \to \chi^0_4 \chi^-_2$ & 542.4 & 542.2 & $0.39^{+11.3 \%}_{-9.6 \%}$  &  $0.47^{+2.7 \%}_{-3.2 \%}{}^{+4.9 \%}_{-3.6 \%}$  & $0.46^{+0.5 \%}_{-1.1 \%}{}^{+4.9 \%}_{-3.7 \%}$\\
$p p \to \chi^+_1 \chi^0_1$ & 303.5 & 161.7 & $0.38^{+6.0 \%}_{-5.4 \%}$  &  $0.46^{+2.5 \%}_{-2.4 \%}{}^{+2.8 \%}_{-2.1 \%}$  & $0.46^{+0.2 \%}_{-0.5 \%}{}^{+2.9 \%}_{-2.1 \%}$\\
$p p \to \chi^+_1 \chi^0_2$ & 303.5 & 303.8 & $35.16^{+6.3 \%}_{-5.8 \%}$  &  $40.90^{+1.6 \%}_{-1.7 \%}{}^{+2.9 \%}_{-2.2 \%}$  & $40.51^{+0.0 \%}_{-0.3 \%}{}^{+2.9 \%}_{-2.2 \%}$\\
$p p \to \chi^+_1 \chi^0_3$ & 303.5 & 526.5 & $0.34^{+9.2 \%}_{-7.9 \%}$  &  $0.40^{+2.6 \%}_{-2.9 \%}{}^{+3.7 \%}_{-2.4 \%}$  & $0.40^{+0.0 \%}_{-0.3 \%}{}^{+3.6 \%}_{-2.5 \%}$\\
$p p \to \chi^+_1 \chi^-_1$ & 303.5 & 303.5 & $25.64^{+6.6 \%}_{-5.9 \%}$  &  $30.37^{+1.7 \%}_{-1.9 \%}{}^{+2.7 \%}_{-2.0 \%}$  & $30.04^{+0.0 \%}_{-0.5 \%}{}^{+2.7 \%}_{-2.1 \%}$\\
$p p \to \chi^+_2 \chi^0_3$ & 542.2 & 526.5 & $1.27^{+11.1 \%}_{-9.4 \%}$  &  $1.46^{+2.9 \%}_{-3.3 \%}{}^{+4.4 \%}_{-2.7 \%}$  & $1.45^{+0.3 \%}_{-0.7 \%}{}^{+4.3 \%}_{-2.9 \%}$\\
$p p \to \chi^+_2 \chi^0_4$ & 542.2 & 542.4 & $1.21^{+11.2 \%}_{-9.5 \%}$  &  $1.37^{+2.7 \%}_{-3.2 \%}{}^{+4.4 \%}_{-2.8 \%}$  & $1.36^{+0.4 \%}_{-0.8 \%}{}^{+4.6 \%}_{-2.6 \%}$\\
$p p \to \chi^+_2 \chi^-_2$ & 542.2 & 542.2 & $0.86^{+10.9 \%}_{-9.3 \%}$  &  $1.00^{+2.6 \%}_{-3.1 \%}{}^{+4.0 \%}_{-2.4 \%}$  & $0.99^{+0.4 \%}_{-0.9 \%}{}^{+4.1 \%}_{-2.4 \%}$\\
\hline
\end{tabular}
\caption{\label{tab:4}
 Total cross sections related to the production of various gaugino pairs of masses 
 $m_1$ and $m_2$, presented together with the associated scale and PDF
 uncertainties
 for the LHC running at a center-of-mass energy of $\sqrt{s}=8$ TeV in
 the context of the benchmark point 1 of the LPCC numbering scheme. The cross
 sections are given at the leading order and next-to-leading order of
 perturbative QCD and matched to threshold resummation.
 The PDF uncertainties are not shown for the LO results. Any cross section smaller
 than 0.1 fb is omitted.}
\end{center}
\end{table}
%

The LO results, given in the fourth column of the Tables, are computed as in
Ref.\
\cite{Debove:2008nr} after convolving (unpolarized) partonic cross sections
with the LO set of the MSTW 2008 parton densities \cite{Martin:2009iq}, accounting
for five light flavors of massless quarks and as agreed among
the SUSY working groups of ATLAS, CMS and the LPCC. We use a top quark mass
of 173.1 GeV
\cite{Group:2009ad} and the values of $m_Z=91.1876$ GeV and $m_W=80.403$ GeV
\cite{Nakamura:2010zzi} for the masses of the electroweak gauge bosons, and we
consider the CKM matrix as the identity matrix in flavor space. The
supersymmetric spectra have been generated with the
{\sc SuSpect} 2.41 program \cite{Djouadi:2002ze}, allowing to obtain 
low-energy supersymmetric masses and parameters from universal parameters
defined at the Grand
Unification scale and evolved down through renormalization group running at the
two-loop level. The central values of the cross sections in the
Tables have been computed after fixing the factorization scale $\mu_F$ to the
average final state particle mass, while the uncertainties are estimated after
multiplying the central value of the factorization scale by a factor lying
in the 0.5--2 range.

%
\begin{table}[!t]
\renewcommand{\arraystretch}{1.2}
\begin{center}
\begin{tabular}{| c || l | l || l | l | l |}
\hline
 Process & $m_1$ [GeV] & $m_2$ [GeV] &  LO [fb] & NLO [fb] & NLO+NLL [fb] \\
\hline
$p p \to \chi^0_1 \chi^0_1$ & 249.6 & 249.6 & $0.13^{+8.6 \%}_{-7.5 \%}$  &  $0.16^{+3.5 \%}_{-3.4 \%}{}^{+3.3 \%}_{-2.3 \%}$  & $0.16^{+0.2 \%}_{-0.3 \%}{}^{+3.5 \%}_{-2.4 \%}$\\
$p p \to \chi^0_2 \chi^-_1$ & 471.9 & 471.8 & $1.63^{+10.0 \%}_{-8.6 \%}$  &  $1.88^{+1.8 \%}_{-2.4 \%}{}^{+4.1 \%}_{-3.1 \%}$  & $1.86^{+0.6 \%}_{-1.2 \%}{}^{+4.1 \%}_{-3.1 \%}$\\
$p p \to \chi^+_1 \chi^0_2$ & 471.8 & 471.9 & $4.73^{+9.8 \%}_{-8.4 \%}$  &  $5.28^{+1.8 \%}_{-2.4 \%}{}^{+3.9 \%}_{-2.5 \%}$  & $5.22^{+0.3 \%}_{-0.6 \%}{}^{+4.0 \%}_{-2.5 \%}$\\
$p p \to \chi^+_1 \chi^-_1$ & 471.8 & 471.8 & $3.13^{+9.8 \%}_{-8.4 \%}$  &  $3.57^{+1.9 \%}_{-2.5 \%}{}^{+3.5 \%}_{-2.2 \%}$  & $3.52^{+0.4 \%}_{-0.7 \%}{}^{+3.7 \%}_{-2.3 \%}$\\
$p p \to \chi^+_2 \chi^0_3$ & 766.3 & 754.0 & $0.16^{+14.2 \%}_{-11.6 \%}$  &  $0.17^{+3.5 \%}_{-4.2 \%}{}^{+6.1 \%}_{-3.8 \%}$  & $0.17^{+1.0 \%}_{-1.8 \%}{}^{+6.1 \%}_{-3.8 \%}$\\
$p p \to \chi^+_2 \chi^0_4$ & 766.3 & 766.6 & $0.15^{+14.3 \%}_{-11.7 \%}$  &  $0.16^{+3.4 \%}_{-4.2 \%}{}^{+6.1 \%}_{-3.9 \%}$  & $0.16^{+1.1 \%}_{-1.8 \%}{}^{+6.1 \%}_{-4.0 \%}$\\
$p p \to \chi^+_2 \chi^-_2$ & 766.3 & 766.3  & $0.11^{+13.6 \%}_{-11.2 \%}$  &  $0.12^{+3.1 \%}_{-3.9 \%}{}^{+6.0 \%}_{-3.5 \%}$  & $0.12^{+1.0 \%}_{-1.8 \%}{}^{+6.0 \%}_{-3.6 \%}$\\
\hline
\end{tabular}
\caption{\label{tab:5}
 Same as Table \ref{tab:4} for the benchmark point 18 of the LPCC numbering
scheme.}
\end{center}
\end{table}
%

For our NLO predictions, we follow the detailed computations performed in our
previous work \cite{Debove:2010kf}, where both the QCD and the SUSY-QCD
contributions are  included in the virtual pieces of the cross section, and give
our results in the fifth column of the Tables. We emphasize that we also
consider, as in our earlier calculations,  
internal squark mixing, {\it i.e.}, the squarks are kept non-degenerate in the
loops. The partonic cross sections are convolved this time with the MSTW 2008
NLO parton density sets \cite{Martin:2009iq}, and the central values of the total
cross sections are given in the Tables together with the theoretical
uncertainties related to scale variation and the choice of the PDF set.
Scale uncertainties are derived following the standard approach of
simultaneously varying the factorization scale $\mu_F$ and
renormalization scale $\mu_R$ by multiplying and dividing the average final-state
particle mass by a factor of two. The PDF errors are obtained by
evaluating the envelope of the cross section when employing the 68\% confidence
level range of the MSTW 2008 parton densities. The (asymmetric) errors $\Delta
\sigma_{\rm up}$ and $\Delta \sigma_{\rm down}$ are computed as defined by
the MSTW collaboration, {\textit{i.e.}, according to
\beq\bsp
  (\Delta \sigma_{\rm up})^2 = &\ \sum_{k=1}^n \Big\{ {\rm max}
    \big[\sigma_k^+ - \sigma_0, \quad \sigma_k^- - \sigma_0,\quad 0 \big]
  \Big\}^2\ ,  \\
  (\Delta \sigma_{\rm down})^2 = &\ \sum_{k=1}^n \Big\{ {\rm max}
    \big[\sigma_0-\sigma_k^+,\quad \sigma_0-\sigma_k^-,\quad 0 \big]\Big\}^2 \ ,
\esp\eeq
where $\sigma_0$ consists in the value of the cross section computed when using
the central set of parton densities, while $\sigma_k^+$ and $\sigma_k^-$ are
those obtained from $\pm 1\sigma$ variations along the $k^{\rm th}$ eigenvector
of the covariance matrix associated to the PDF fit.

%
\begin{table}[!t]
\renewcommand{\arraystretch}{1.2}
\begin{center}
\begin{tabular}{| c || l | l || l | l | l |}
\hline
 Process & $m_1$ [GeV] & $m_2$ [GeV] &  LO [fb] & NLO [fb] & NLO+NLL [fb] \\
\hline
$p p \to \chi^0_1 \chi^0_1$ & 251.7 & 251.7  & $0.12^{+8.6 \%}_{-7.5 \%}$  &  $0.16^{+3.5 \%}_{-3.4 \%}{}^{+3.3 \%}_{-2.3 \%}$  & $0.15^{+0.2 \%}_{-0.3 \%}{}^{+3.4 \%}_{-2.4 \%}$\\
$p p \to \chi^0_2 \chi^-_1$ & 478.5 & 478.5 & $1.50^{+10.1 \%}_{-8.6 \%}$  &  $1.73^{+1.8 \%}_{-2.4 \%}{}^{+4.2 \%}_{-3.1 \%}$  & $1.71^{+0.6 \%}_{-1.2 \%}{}^{+4.1 \%}_{-3.2 \%}$\\
$p p \to \chi^+_1 \chi^0_2$ & 478.5 & 478.5 & $4.37^{+9.9 \%}_{-8.5 \%}$  &  $4.86^{+1.8 \%}_{-2.4 \%}{}^{+3.9 \%}_{-2.5 \%}$  & $4.81^{+0.3 \%}_{-0.6 \%}{}^{+4.2 \%}_{-2.6 \%}$\\
$p p \to \chi^+_1 \chi^-_1$ & 478.5 & 478.5 & $2.89^{+9.9 \%}_{-8.5 \%}$  &  $3.28^{+1.9 \%}_{-2.5 \%}{}^{+3.5 \%}_{-2.3 \%}$  & $3.24^{+0.5 \%}_{-0.7 \%}{}^{+3.8 \%}_{-2.3 \%}$\\
\hline
\end{tabular}
\caption{\label{tab:6}
 Same as Table \ref{tab:4} for the benchmark point 31 of the LPCC numbering
scheme.}
\end{center}
\end{table}
%

In the last column of the Tables, we present the total cross section after
having matched the next-to-leading order results with threshold resummation at 
the next-to-leading logarithmic accuracy.
After performing a Mellin transformation, the cross section can be written in
the Mellin $N$-space as a simple product between the parton densities $f_{a/p}$
and $f_{b/p}$ and the partonic cross section $\sigma_{ab}$,
\beq
  M^2\frac{\d\sigma}{\d M^2}(N-1)=\sum_{ab} f_{a/p}(N,\mu^2) 
  f_{b/p}(N,\mu^2)\ \sigma_{ab}(N,M^2,\mu^2) \ ,
\label{eq:mellin}\eeq
where we have set $\mu_F=\mu_R=\mu$ for brevity and where $M$ denotes the
invariant 
mass of the produced gaugino pair. Within the threshold resummation formalism,
the partonic cross section can be refactorized into a closed exponential form as 
\cite{Vogt:2000ci}
\beq
  \sigma_{ab}(N,M^2,\mu^2) = {\cal H}_{ab}(M^2, \mu^2)\ \exp\Big[{\cal
     G}_{ab}(N,M^2,\mu^2)\Big] + {\cal O}\Big(\frac{1}{N}\Big) \ .
\label{eq:res}\eeq
The function ${\cal H}$ can be perturbatively computed, describes the hard
part of the scattering process, and is independent of the Mellin variable
$N$. The function ${\cal G}$ collects soft and collinear parton emission
and absorbs the large logarithmic contributions arising at fixed order. 
We then further improve the resummation formula of Eq.\ \eqref{eq:res} above by
including in the ${\cal H}$-function the dominant $1/N$-terms
stemming from universal collinear radiation of the initial state partons
\cite{Kramer:1996iq, Catani:2001ic, Kulesza:2002rh, Almeida:2009jt}. The
exact expressions of the coefficients of the expansion of the two functions
${\cal H}$ and ${\cal G}$ at the next-to-leading logarithmic accuracy can be
found in Ref.\ \cite{Debove:2010kf}.

The resummed predictions are known to be valid near production threshold, where
the logarithmic terms of the cross section dominate, in contrast to the fixed
order computations, which are assumed to be valid far from this threshold.
Therefore, in order to get reliable predictions in all kinematic regions, both
results are consistently matched by summing the resummed
$\sigma^{\rm(res.)}_{ab}$ and the fixed order $\sigma^{\rm(f.o.)}_{ab}$ cross
sections and subtracting their overlap $\sigma^{\rm(exp.)}_{ab}$,
\beq
  \sigma_{ab} = \sigma^{\rm(res.)}_{ab}+\sigma^{\rm(f.o.)}_{ab}-
  \sigma^{\rm(exp.)}_{ab} \ .
\label{eq:matching}\eeq
We again refer to Ref.\ \cite{Debove:2010kf} for the exact expression of
$\sigma^{\rm(exp.)}_{ab}$ which is, for a NLO+NLL matching, the ${\cal
O}(\alpha_s)$ series expansion of the resummed cross section of Eq.\
\eqref{eq:res}. Since the two quantities $\sigma^{\rm(res.)}_{ab}$ and
$\sigma^{\rm(exp.)}_{ab}$ are computed in Mellin space, an inverse transform
must be performed in order to obtain the total cross section in terms of
physical quantities
\beq\bsp
 M^2{\d\sigma\over\d M^2}(\tau) =&\ {1\over2\pi i} \sum_{ab} \int_{{\cal C}_N}\d N 
 \tau^{-N} f_{a/p}(N+1,\mu^2) 
  f_{b/p}(N+1,\mu^2)\\
  &\quad \times \Big[\sigma_{ab}^{\rm(res.)}(N+1,M^2,\mu^2) -
   \sigma_{ab}^{\rm(exp.)}(N+1,M^2,\mu^2) \Big] \ ,
\esp\label{eq:inverse}\eeq
where $\tau=M^2/s$. To avoid singularities in the integrand related to the
Landau pole of the strong coupling constant and to the Mellin moments of the
parton densities at small momentum fraction, we choose an integration contour
${\cal C}_N$ according to the principal value procedure
\cite{Contopanagos:1993yq} and the minimal prescription  \cite{Catani:1996yz}.

The resummed results are given in Tables \ref{tab:4}, \ref{tab:5} and
\ref{tab:6} together with the uncertainties derived from scale variations and
deviations from the central fit of the MSTW 2008 parton densities. The latter
are computed as for the pure NLO results and described above. 

One observes, by
investigating the numerical results, an improvement of the stability of the
perturbative series, as the scale dependence of the total cross section is tamed
by the resummation of the soft and collinear radiation contributions. At LO, the
factorization
scale $\mu_F$ appearing within the evolution of the parton densities already
induces large logarithmic terms yielding an uncertainty of about $10\%$ on the
cross section. This effect is then attenuated at NLO, but one gets an
additional explicit dependence in the renormalization scale $\mu_R$ through the
strong coupling constant and the loop contributions. After matching the NLO
results with threshold resummation at NLL, the scale variations are eventually
considerably
reduced thanks to the inclusion of dominant higher order contributions
within the Sudakov form factor ${\cal G}$. 
In contrast, the uncertainties related to the choice of the employed parton
densities coincide when comparing the NLO and the NLO+NLL results, which is
not surprising since the same PDF sets enter both computations.

\subsection{Total cross sections for model lines 10.1, 10.3, and 40.1}
\label{sec:2c}

In order to give a more complete overview of the absolute size of the total gaugino cross
sections, we present in this Section the total cross sections for the golden trilepton channel,
{\it i.e.} the production of a $\tilde \chi_2^0 \tilde \chi_1^+$ pair, as a function of
their almost equal mass $m_{\tilde{\chi}}$ at the LHC with a center-of-mass energy of 8 TeV

The results for model line 10.1 are shown in Figure \ref{fig:1} at LO (dotted), NLO (dashed),
%
\begin{figure}[t!]
 \centering
 \epsfig{file=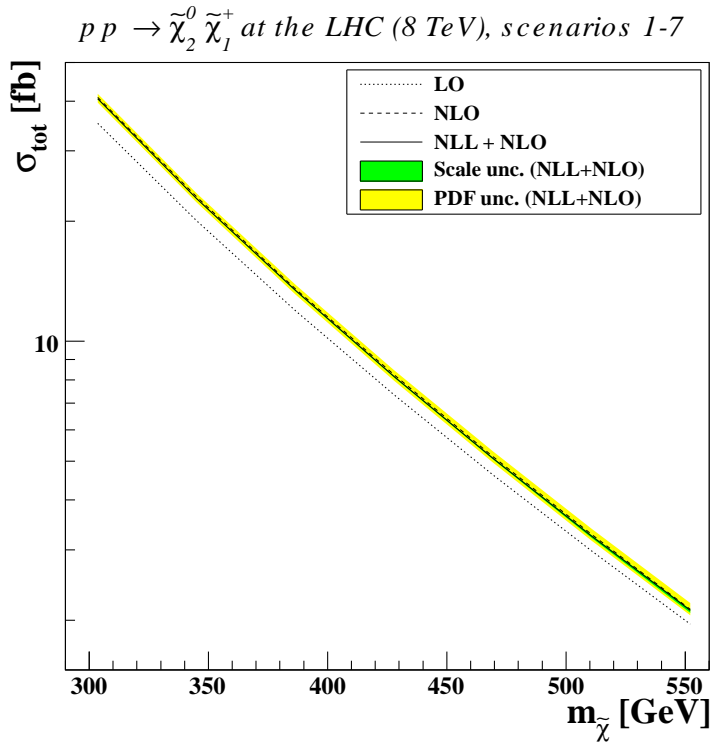,width=.7\textwidth}
 \caption{\label{fig:1}
 The total cross section at LO (dotted), NLO (dashed), and NLO+NLL (full) with its scale
 (green) and PDF (yellow) uncertainty for the production of a $\chi^0_2\chi^+_1$ pair
 as a function of their almost degenerate mass $m_{\tilde{\chi}}$
 at the LHC with $\sqrt{s}=8$ TeV and
 benchmark points 1--7.}
\end{figure}
%
and NLL+NLO (full). As the soft SUSY-breaking mass $m_{1/2}$ increases from 400 to 700 GeV,
the gaugino masses rise from 300 to 550 GeV. Consequently, the total cross section falls by
over an order of magnitude from about 40 fb to about 2 fb. The impact of the NLO corrections
to the LO prediction is clearly visible, and the cross section is then stabilized at NLL+NLO,
where it remains almost unchanged with respect to the NLO result, but is considerably less
scale-dependent. The overall theoretical uncertainty is thus dominated by the PDF uncertainty,
estimated as described in the previous Section. Similar results are obtained for
model lines 10.3 and 40.1 (not shown), as they have similar physical gaugino
mass ranges.

\section{Comparisons of invariant-mass and transverse-momentum distributions}
\label{sec:3}

Typical final-state signatures associated to the production of the electroweak
superpartners of the Standard Model gauge and Higgs bosons contain in general
abundant initial state QCD radiation. The
latter has important effects on the shapes of the kinematical distributions, in
particular due to the large logarithmic contributions arising in the phase space
regions where these additional partons are neither widely 
separated nor hard. In this case, reliable theoretical predictions require a
consistent reorganization of the logarithmic terms so that they are effectively
resummed to all orders in the strong coupling constant and embedded in 
the so-called Sudakov form factor. In the resummation computations presented in
Section \ref{sec:2}, this factor has been derived at the next-to-leading
logarithmic
accuracy from the knowledge of the perturbative expansion of the hard-scattering
function at the next-to-leading order. This leads to an accurate description of
the QCD environment in phase space regions where radiation is soft and/or
collinear. In order to obtain a good description over the whole kinematical range,
including also the hard radiation regime, the resummed results have to be
eventually matched to a fixed order calculation describing precisely the hard 
emission.

In experimental analyses or phenomenological investigations relying on Monte
Carlo simulations, the production of additional jets is traditionally simulated
using parton showering programs, such as {\sc Herwig} \cite{Corcella:2000bw,Bahr:2008pv}
or {\sc Pythia} \cite{Sjostrand:2006za,Sjostrand:2007gs}, which
describe QCD emissions as successive branchings of a mother parton into two
daughter partons. These tools are based on Markov chain techniques built upon
the Sudakov form factor dictating the probability laws for a specific parton to 
radiate or not. In contrast to the resummation techniques adopted for the
predictions shown in Section \ref{sec:2}, where the Sudakov form factor is 
computed at the next-to-leading logarithmic accuracy, the parton showering tools
use only a leading-logarithmic precision, sometimes including partial (but not
complete) next-to-leading logarithmic contributions. The reason is that only such 
approximations allow for a description in terms of a Monte Carlo algorithm.
Moreover, for supersymmetric processes the hard scattering
can only be evaluated at tree level, since no SUSY process has so far been
implemented into NLO Monte Carlo tools such as {\sc MC@NLO}
\cite{Frixione:2002ik, Frixione:2003ei} or {\sc Powheg} \cite{Nason:2004rx}.
This description is formally only correct in the soft and collinear
radiation regions of the phase space, and it fails when considering the
production of hard and widely separated additional partons due to missing 
subleading terms. In this case, matrix elements describing the same final state 
together with an additional parton are required.

The matrix-element approach is indeed accurate for describing hard and widely
separated emissions, but is known to break down in the soft and collinear
radiation limits. Contrary, parton showering gives a proper description in the
soft and collinear kinematical regions, but underestimates hard emissions.
Therefore, several matching algorithms, such as the Catani-Krauss-Kuhn-Webber
(CKKW) scheme based on event reweighting \cite{Catani:2001cc,Krauss:2002up} or 
the Mangano (MLM) scheme based on event rejection \cite{Mangano:2006rw}, have
been developed to combine the two methods in a consistent way. Hence this avoids 
the possible double counting arising from radiations which can be taken
into account both at the level of the matrix element and at the level of the 
Sudakov form
factor. These different procedures have been extensively applied, compared and 
confronted to the data in the context of various Standard Model processes
\cite{Mrenna:2003if, Krauss:2004bs, Hoche:2006ph, Alwall:2007fs} and also used
for recent Beyond the Standard Model explorations \cite{Alwall:2008qv,
deAquino:2011ix,deAquino:2012ru}.

In this Section, we present the first comparison between resummation (matched to
fixed order), next-to-leading order and parton showering (matched to matrix
elements) predictions in
the context of chargino and neutralino production. We focus on the 
distributions of the transverse momentum ($p_T$) and invariant mass ($M$) of the
superparticle pairs produced at the LHC collider, running at a center-of-mass
energy of 8 TeV. Concerning the most precise predictions, we employ the
transverse-momentum and threshold resummation formalisms. Threshold resummation
has been briefly described in Section \ref{sec:2b} and more details can be found
in Ref.\ \cite{Debove:2010kf}. As for the threshold case (see Eq.\
\eqref{eq:mellin}), resummation in the transverse-momentum regime is also
performed in Mellin space. The partonic cross is computed in the impact
parameter $b$-space
\beq
 \sigma_{ab}(p_T^2,M^2,\mu^2) =
 {M^2\over s} \int_0^\infty \d b {b\over 2}  J_0 (b p_T) \
 \sigma_{ab} (b, N, M^2,s,\mu^2) \ ,
\eeq
where the function $J_0$ is the $0^{\rm th}$-order Bessel function, so that 
the integrand can be refactorized under a closed
exponential form, as in Eq.\ \eqref{eq:res} for threshold resummation,
\beq
  \sigma_{ab}(b, N,M^2,\mu^2) = {\cal H}_{ab}(N, M^2, \mu^2)\ \exp\Big[{\cal
     G}_{ab}(b, N,M^2,\mu^2)\Big] + {\cal O}\Big(\frac{1}{N}\Big) \ .
\eeq
As usual, the Sudakov form factor ${\cal G}$ contains the soft and collinear
radiation contributions and ${\cal H}$ is the hard function, independent of the
impact parameter $b$. More details and the exact expressions of these functions,
evaluated at the next-to-leading logarithmic accuracy,
can be found in Ref.\ \cite{Debove:2009ia}. Matching to the fixed order
computations and the inverse Mellin transform to get back to the physical space are 
eventually performed, as sketched in Eqs.\ \eqref{eq:matching} and
\eqref{eq:inverse}.

Concerning the merging of the
parton showering with the hard-scattering matrix elements, we follow the
$k_T$-MLM matching scheme \cite{Alwall:2008qv} as implemented in the {\sc
MadGraph}-{\sc MadEvent} event generator version 5 \cite{Alwall:2011uj},
interfaced to {\sc Pythia} 6 \cite{Sjostrand:2006za}.  
The UFO \cite{Degrande:2011ua} model files required for {\sc MadGraph} are
generated starting from the MSSM implementation in the {\sc FeynRules} package
\cite{Christensen:2008py, Christensen:2009jx, Duhr:2011se} after loading the
supersymmetric spectrum associated to the benchmark points 1, 18 and 31
described in Section \ref{sec:2} and obtained as in Section \ref{sec:2b} with
the {\sc SuSpect} 2.41 program \cite{Djouadi:2002ze}. We generate
three distinct parton-level event samples for final states containing, in
addition to the pair of supersymmetric particles, zero, one and two extra
partons, respectively. We subsequently merge them after parton-showering,
employing the $k_T$-MLM matching scheme.
The (parton-level) jets are generated with a minimum jet measure
$k_T$ of 50 GeV between two final state partons $i$ and $j$, where $k_T$ is
defined by
\beq
  k_T^2 = \min(p_{Ti}^2, p_{Tj}^2) R_{ij}\ . 
\eeq
In the equation above, the quantities $p_{Ti}$ and
$p_{Tj}$ are the transverse momenta of the two partons under consideration and
$R_{ij}$ denotes their angular distance in the $(\eta,\phi)$ plane. In the case
where one of the two partons is an initial state parton, we define the jet measure as
the transverse momentum of the final state parton
\beq
  k_T = p_{Ti}
\eeq
and require its value to be larger than 20 GeV, which ensures better
QCD factorization properties with respect to initial-state collinear
singularities \cite{Catani:1993hr}. The events are
then passed to
{\sc Pythia} for showering, and jets are reconstructed using {\sc FastJet}
\cite{Cacciari:2005hq} in the $k_T$-jet algorithm with
a cut-off scale $Q^{\rm match}$ set to 70 GeV. The jets are said to be matched
to one of the original partons if the $k_T$-measure between the jet and the
parton
is smaller than $Q^{\rm match}$. The events are selected only if each jet is
matched to one parton, with the exception of the two-jet sample. In this case, 
extra jets are still allowed, which maintains the full inclusiveness of the
matched sample. 

In our simulation setup, we neglect all quark masses but the top mass and 
employ the leading order set of the MSTW 2008 parton density fits
\cite{Martin:2009iq}. In contrast, the NLO and NLO+NLL results are based on 
the
next-to-leading order set of the MSTW 2008 fits. We identify the factorization
and renormalization
scales as the average mass of the produced particles in the case of the three
hard processes, {\it i.e.}, without any extra parton as well as with one and
with two additional (hard) radiations, while the renormalization scale used by
the showering algorithm consists in the jet measure at each branching. 
The value of the total cross section associated to a matched event sample is
usually rather close to the one of the original process, where no extra
radiation is allowed, \textit{i.e.}, it is close to the value of the tree-level
cross section. Therefore, we reweight the produced events uniformly, so that the
total rate is now equal to the resummed cross sections given in the Tables of
Section \ref{sec:2}. 
Let us note that we have also performed the on-shell subtraction of any
intermediate resonance which could appear in the samples containing one or two
extra partons using the narrow-width approximation, as those subprocesses are
rather identified with other genuine processes such as the associated production
of a gaugino and a squark or a gluino. Finally, the event samples have been
analyzed with the program package {\sc MadAnalysis} 5 \cite{Conte:2012fm}.

\subsection{Distributions in the invariant mass of the gaugino pair}
\label{sec:3a}

%
\begin{figure}
 \centering
 \epsfig{file=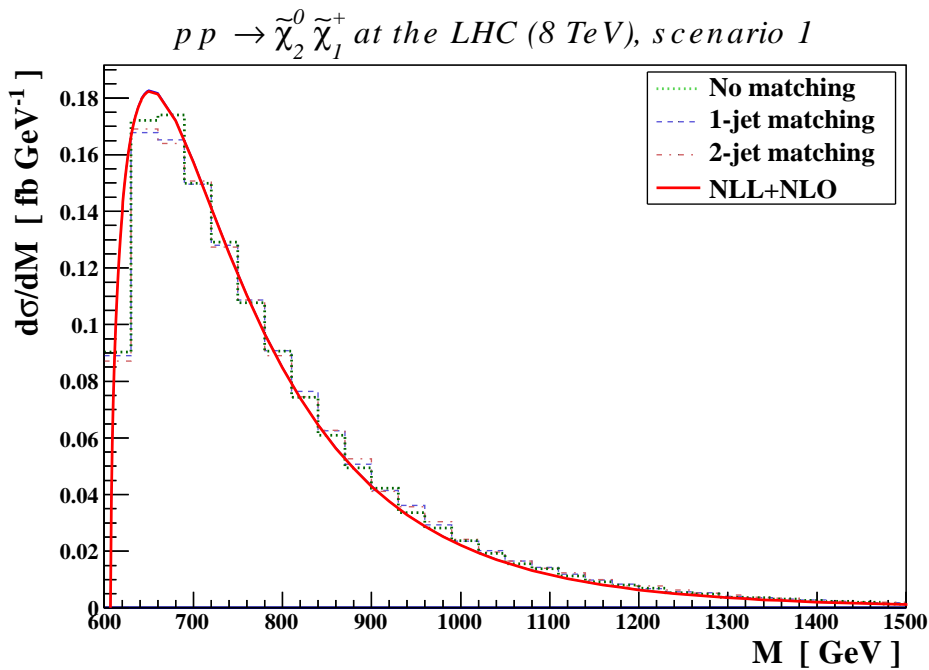,width=.9\textwidth}
 \caption{\label{fig:3}Distributions in the invariant mass $M$ of a
$\tilde \chi^0_2 \tilde\chi^+_1$ pair with mass 304 GeV each (benchmark point 1) at the LHC
with $\sqrt{s}=8$ TeV. We compare 
the NLO matched to the NLL (red, thick full) distribution to the results
obtained after matching matrix elements containing no (green, dotted), one (blue,
dashed), and up to two (red, dot-dashed) additional jets to parton showering.}
\end{figure}
%

In Figure \ref{fig:3}, we depict the invariant mass
spectrum $\d\sigma/\d M$ for the associated production of a lightest chargino
($\tilde \chi_1^+$) and second lightest neutralino ($\tilde \chi_2^0$) pair
in the context of the
benchmark scenario 1 presented in Section \ref{sec:2a}. We compare the results
obtained using the threshold resummation formalism (thick red full curve),
found to be almost equivalent to those corresponding to a pure next-to-leading
order calculation, to those generated by LO Monte Carlo simulations
with {\sc MadGraph} and interfaced to the parton showering algorithm
provided by
{\sc Pythia}. For the latter, we have normalized the total cross section to
its resummed value of 40.51 fb (see Table \ref{tab:4}). The spectrum starts
at $M\sim 600$ GeV, since the two gauginos are almost degenerate with a mass of
about 300 GeV each. The shape of the spectrum is then dictated by the nature of
the produced superparticles, which are here mainly gaugino-like. This suppresses the
$s$-channel $W$-boson exchange contributions with respect to $P$-wave
production through $t$- and $u$-channel squark exchange, the mass of the squarks
being here of the order of 800 GeV. The
resummed contributions increase the total rate in comparison to the
pure next-to-leading order results only slightly, since threshold effects only become important
when the invariant mass approaches the total center-of-mass energy.

The set of three curves obtained using the {\sc MadGraph} and {\sc Pythia}
programs agree quite well with each other. They are related to three
different ways which we have adopted to perform the matching procedure. The 
results represented by the green dotted curve have been obtained from tree-level
matrix elements only, related hence to the process
\beq
  p p \to \tilde \chi_2^0 \tilde \chi_1^+ 
\eeq
without any extra radiation. The parton-level events have then been passed to
{\sc Pythia} for parton showering, and no matching procedure has been
applied. After accounting for a resummed $K$-factor normalizing the integrated
distribution to 40.51 fb (see Table \ref{tab:4}), one observes a good agreement 
with the resummed predictions. A similar behavior has already been
observed in the context of $Z'$ production at the LHC, where in Ref.\
\cite{Fuks:2007gk} resummed computations were confronted to results obtained
with tree-level matrix-elements passed to a parton showering algorithm. 

Applying the $k_T$-MLM matching procedure described in the beginning of Section
\ref{sec:3}, we compute matrix elements containing up to one additional QCD
emission, {\it i.e.}, considering the subprocesses
\beq
  p p \to \tilde \chi_2^0 \tilde \chi_1^+
  \quad\text{\rm and}\quad
  p p \to \tilde \chi_2^0\ \tilde \chi_1^+ j,
\eeq
where $j$ denotes a quark, an antiquark or a gluon, and match them
consistently to parton showering. We obtain the results shown by the blue dashed
curve of Figure \ref{fig:3} after normalizing again the total cross section to
its resummed value of 40.51 fb (see Table \ref{tab:4}).

In a similar fashion, one can allow for matrix elements with up to
two extra partons, {\it i.e.}, we account for the processes
\beq
  p p \to \tilde \chi_2^0\ \tilde \chi_1^+ , \quad
  p p \to \tilde \chi_2^0\ \tilde \chi_1^+ j,
  \quad\text{\rm and}\quad
  p p \to \tilde \chi_2^0\ \tilde \chi_1^+ j\ j \ .
\eeq
This leads to the dot-dashed red curve presented in Figure \ref{fig:3}. Again,
since the matched total cross section is close to the unmatched one, the
results have been normalized to 40.51 fb, including hence a resummed $K$-factor.

%
\begin{figure}
 \centering
 \epsfig{file=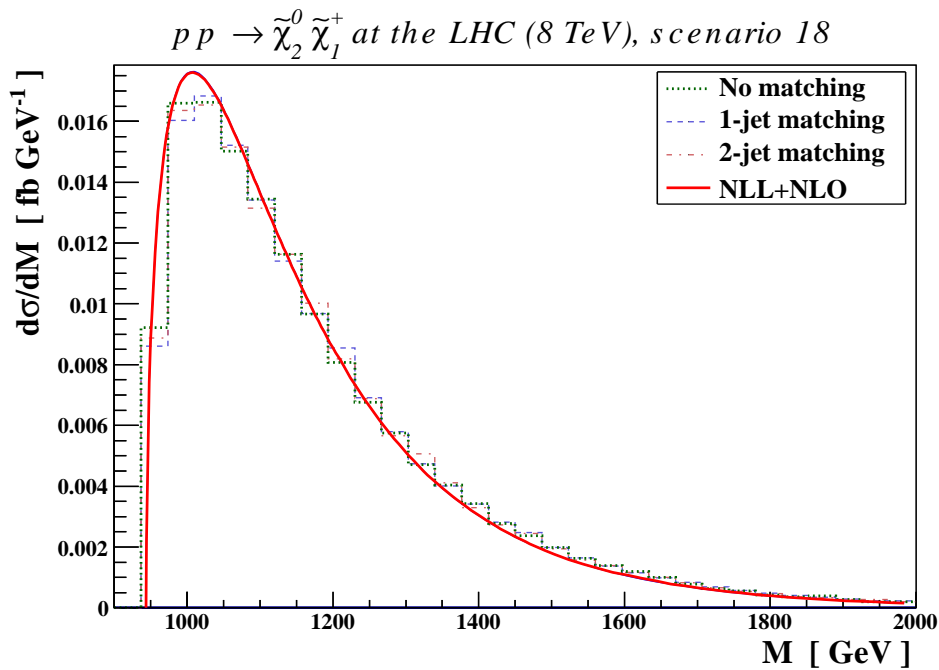,width=.9\textwidth}
 \caption{\label{fig:4}
 Same as Figure \ref{fig:3} for the benchmark point 18. The gaugino masses are
here 472 GeV. The distributions for the benchmark point 31 with gaugino masses
of 479 GeV are very similar and therefore not shown.}
\end{figure}
%

In the case of the benchmark scenarios 18 and 31, the gaugino masses are
heavier, with $m_{\tilde\chi} = 472$ GeV and 479 GeV, respectively. This reduces
the total production cross section of an associated pair of the lightest
chargino and the next-to-lightest neutralino by about an order of magnitude,
which then reads 5.22 fb for the scenario 18 (see Table \ref{tab:5}) and 4.81
fb for the scenario 31 (see Table \ref{tab:6}).
Employing these values for the normalization of the results generated by {\sc
MadGraph} and {\sc Pythia}, we confront them again to our NLL+NLO
calculation for scenario 18 in Figure \ref{fig:4} and for scenario 31 (not
shown). One observes a similar behavior as for the
scenario 1 shown in Figure \ref{fig:3}.

\subsection{Distributions in the transverse momentum of the gaugino pair}

%
\begin{figure}
 \centering
 \epsfig{file=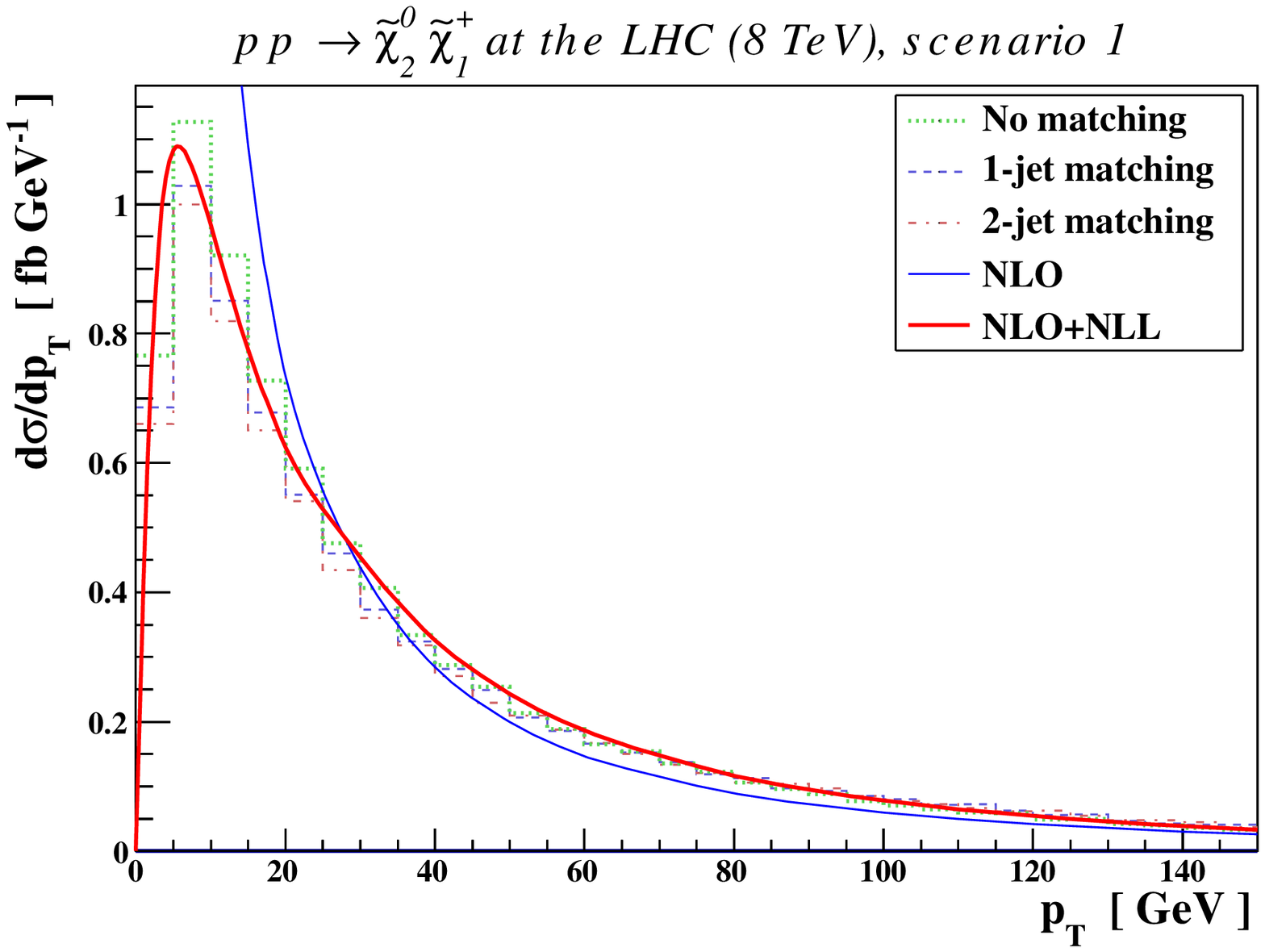,width=.9\textwidth}
 \epsfig{file=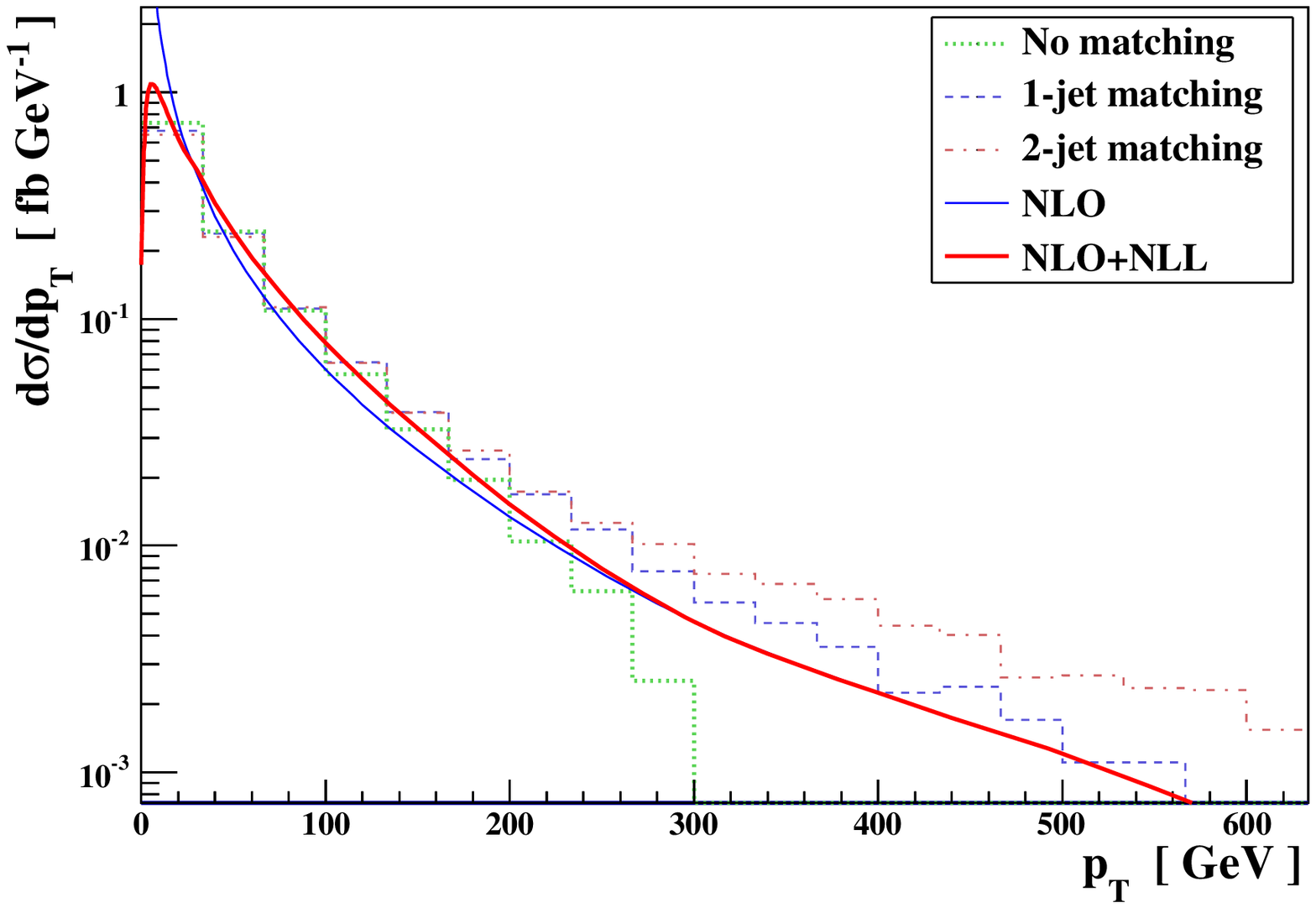,width=.9\textwidth}
 \caption{\label{fig:6}Distributions in the transverse momentum $p_T$ of a
$\tilde\chi^0_2\tilde\chi^+_1$ pair with mass 304 GeV each (benchmark point 1) at the LHC
with $\sqrt{s}=8$ TeV. We compare fixed order at ${\cal O}(\alpha_s)$ (blue, full)
and NLL (red, thick full) distributions to the results
obtained after matching matrix-elements containing no (green, dotted), one (blue,
dashed), and up to two (red, dot-dashed) additional jets to parton showering 
in the small (top) and large (bottom) $p_T$ regions.}
\end{figure}
%

For scenario 1 at the LHC with a center-of-mass
energy of 8 TeV, we show in Figure \ref{fig:6} the transverse-momentum
distribution $\d\sigma/\d p_T$ of a lightest chargino ($\tilde\chi_1^+$)
and next-to-lightest neutralino ($\tilde\chi_2^0$) pair.
As expected, the fixed order predictions at ${\cal O}(\alpha_s)$
(full blue curve) diverge as the transverse momentum $p_T$
tends to zero due to the unbalanced large logarithmic terms related
to soft parton radiation. After their consistent matching to a resummation
computation in the
transverse-momentum regime at the next-to-leading logarithmic accuracy (thick
red full curve), the results exhibit a finite behavior with a maximum around a
$p_T$-value of about 10 GeV. The effects of the resummation of the large
logarithms extend to intermediate values of the transverse momentum, where the
resummed predictions are considerably larger than
those obtained from the pure fixed order computation as it has already been
observed for slepton-pair production \cite{Bozzi:2006fw, Bozzi:2007tea} or other
gaugino-pair production channels \cite{Debove:2009ia, Debove:2011xj}.

We now turn to the confrontation of the resummed results to the three sets of
predictions obtained after employing the {\sc MadGraph} and {\sc Pythia}
generators, following the same approach as the one described in Section
\ref{sec:3a}. Again the total production rate associated to each
of these three curves has been set to the value of the resummed total cross
section of 40.51 fb (see Table \ref{tab:4}). One observes very good agreement
between the most precise distribution including a matching of the resummed
results to the fixed order ones (thick red full curve) and
the distributions obtained with {\sc MadGraph} and {\sc Pythia} after having
matched the matrix elements containing one or two additional jets to parton
showering (blue dashed and red dot-dashed curves in Figure \ref{fig:6}).
Contrary, the simple application of a parton showering algorithm to the
tree-level matrix element (green dotted curve on Figure \ref{fig:6}) leads to a
slightly too soft spectrum as expected, since parton showering methods
can not address properly the intermediate and large $p_T$ regions.

In contrast to the invariant-mass distributions, where the three Monte
Carlo predictions (without matching, with a matching procedure of matrix elements
accounting for at most one additional parton, and with a matching procedure of
matrix elements containing at most two extra partons) have been found to be 
in a rather good
agreement (see Section \ref{sec:3a}), one can clearly here observe the effects of including
additional partons when comparing the three distributions generated by the Monte
Carlo simulators in Figure \ref{fig:6}. While the position of the peak does not
depend much on the presence of additional partons at the
matrix-element level, the global hardness of the spectrum does indeed highly depend on
the contributions of these extra partons. One observes a very good agreement
between the NLL+NLO results and the Monte Carlo predictions after matching
parton showering to matrix elements containing up to one extra parton, which is
not surprising since the same matrix elements, relevant for hard emissions, are
included in the resummation computations. As resummation computations at
NNLO+NNLL must still be  performed, the accuracy of the Monte Carlo predictions
with two hard jets matched to parton showers can at this point not yet be
judged.

%
\begin{figure}
 \centering
 \epsfig{file=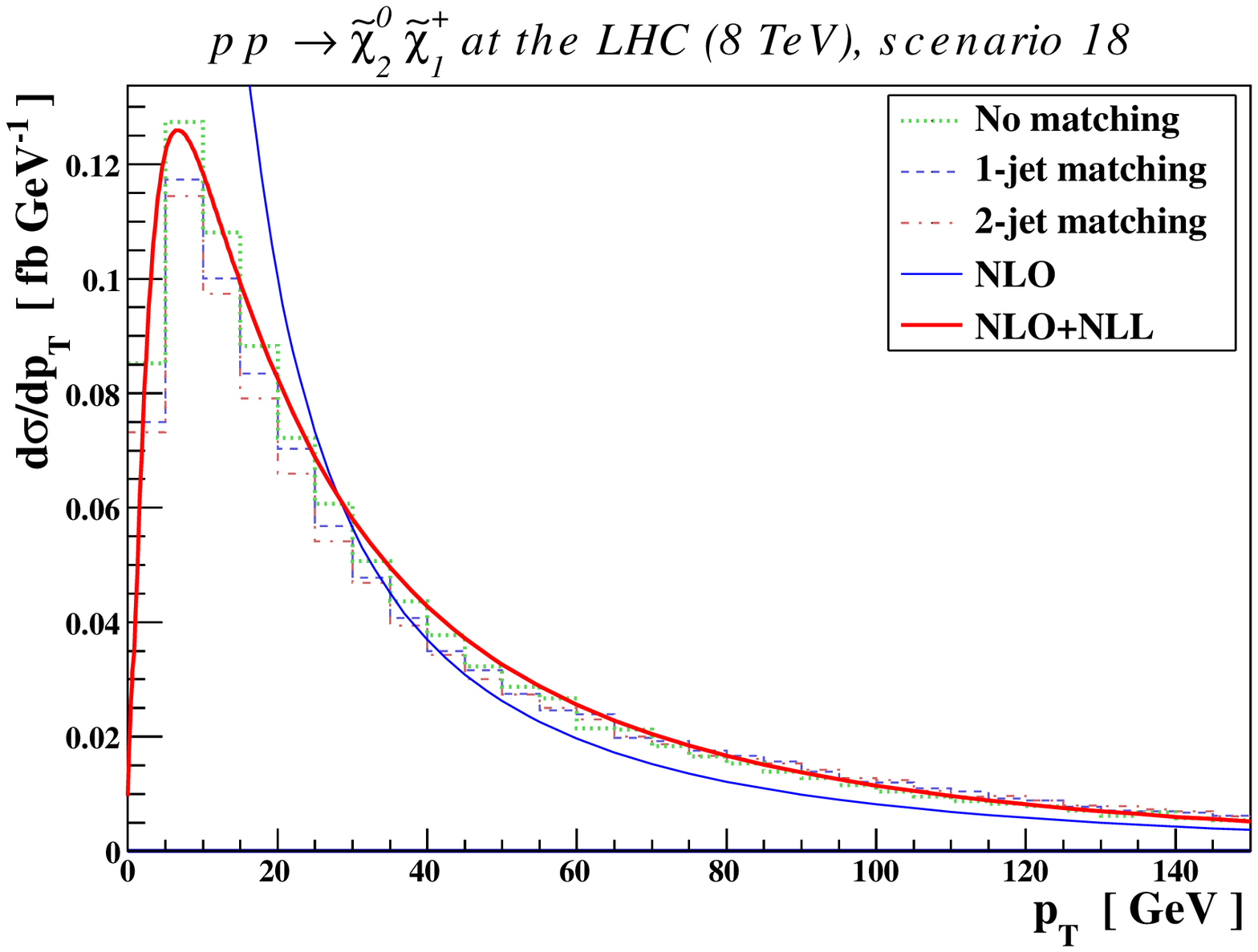,width=.9\textwidth}
 \epsfig{file=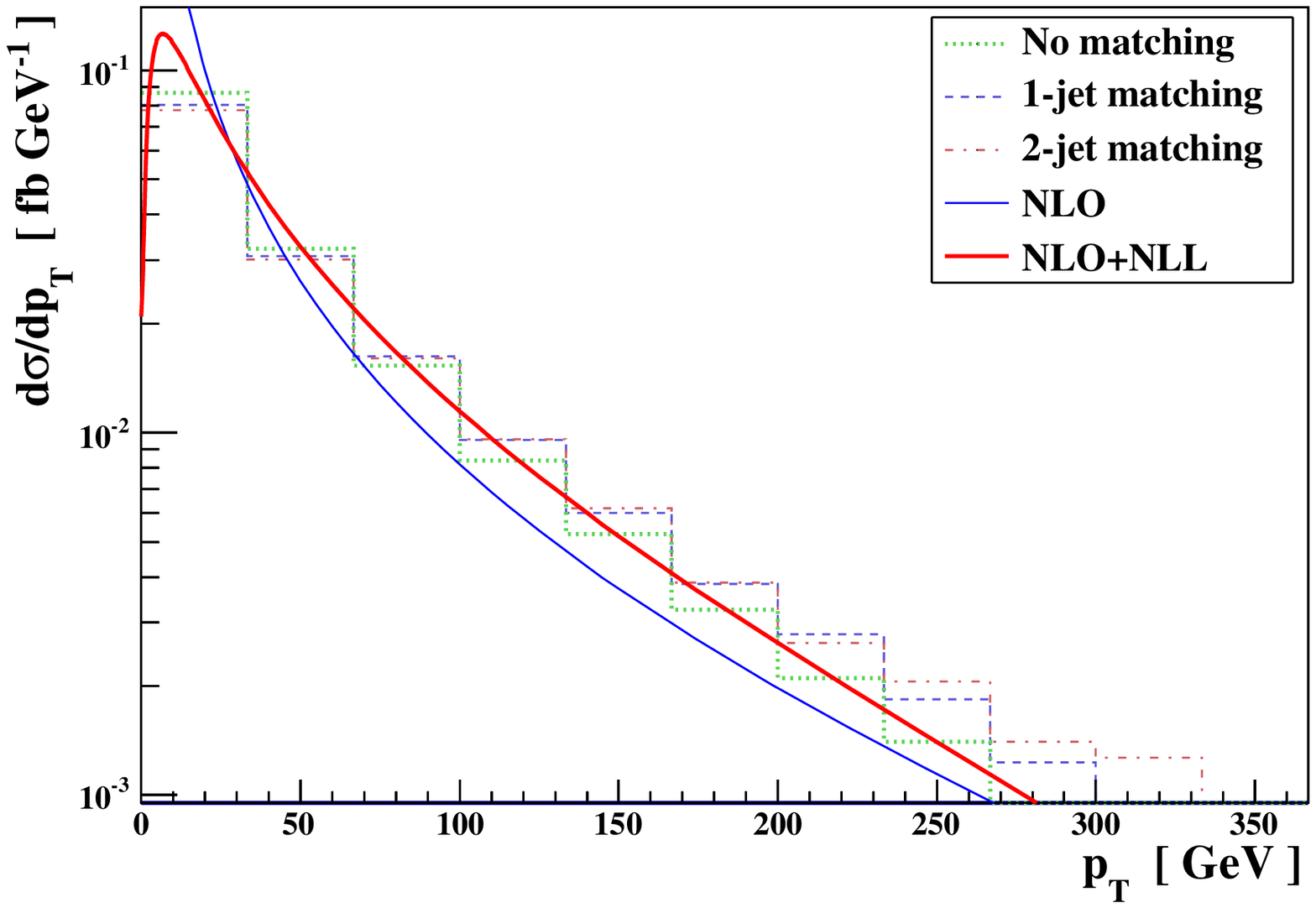,width=.9\textwidth}
 \caption{\label{fig:7}
 Same as Figure \ref{fig:6} for the benchmark point 18. The gaugino masses are
here 472 GeV. The distributions for the benchmark point 31 with gaugino masses
of 479 GeV are very similar and therefore not shown.}
\end{figure}
%

The same type of effects can be noticed for the benchmark scenario 18 in Figure
\ref{fig:7} and scenario 31 (not shown), the only difference being the overall
scale of the
distributions, which is about one order of magnitude smaller than in scenario 1
due to the larger gaugino masses.

%
\begin{figure}
 \centering
 \epsfig{file=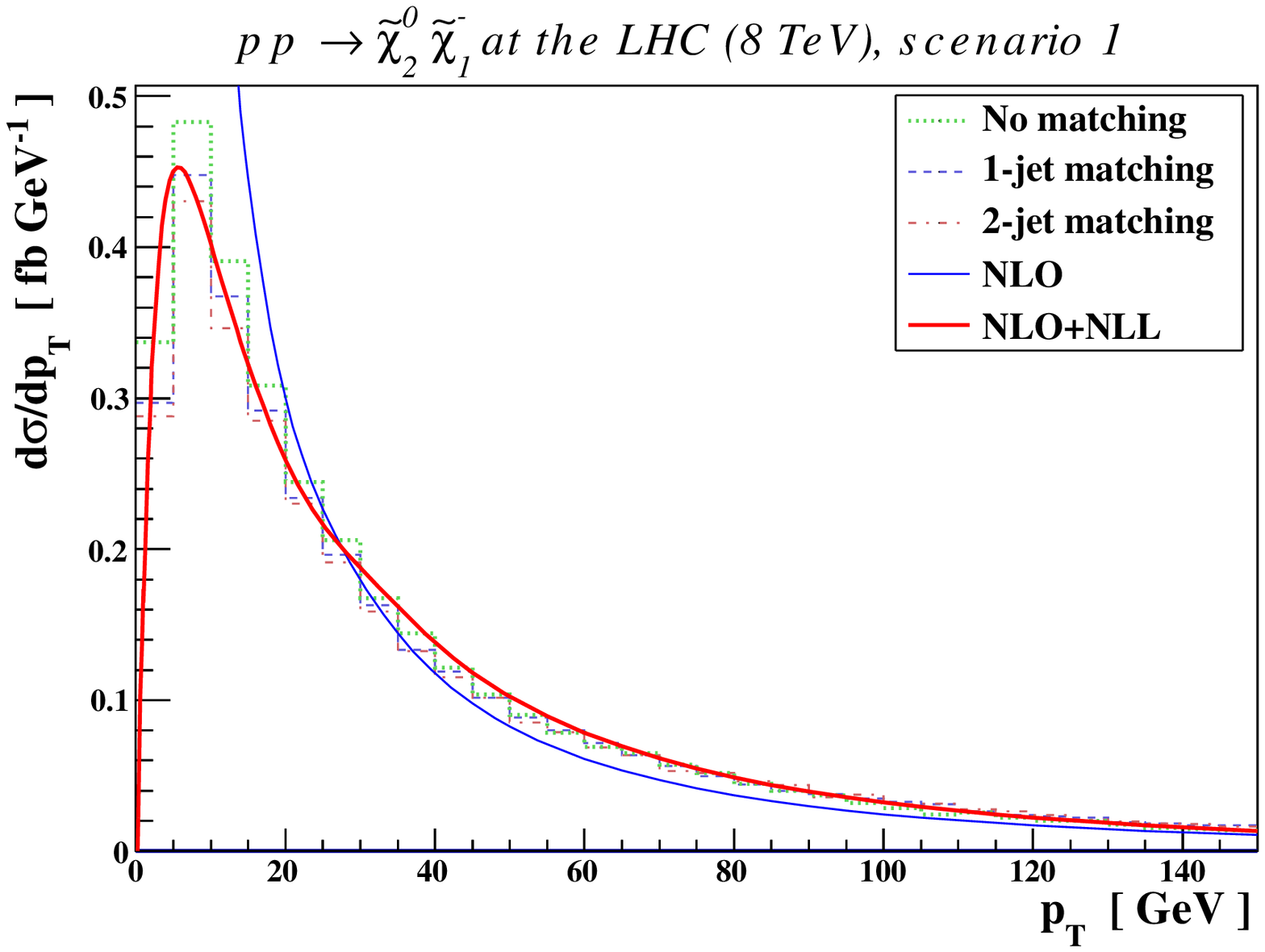,width=.9\textwidth}
 \epsfig{file=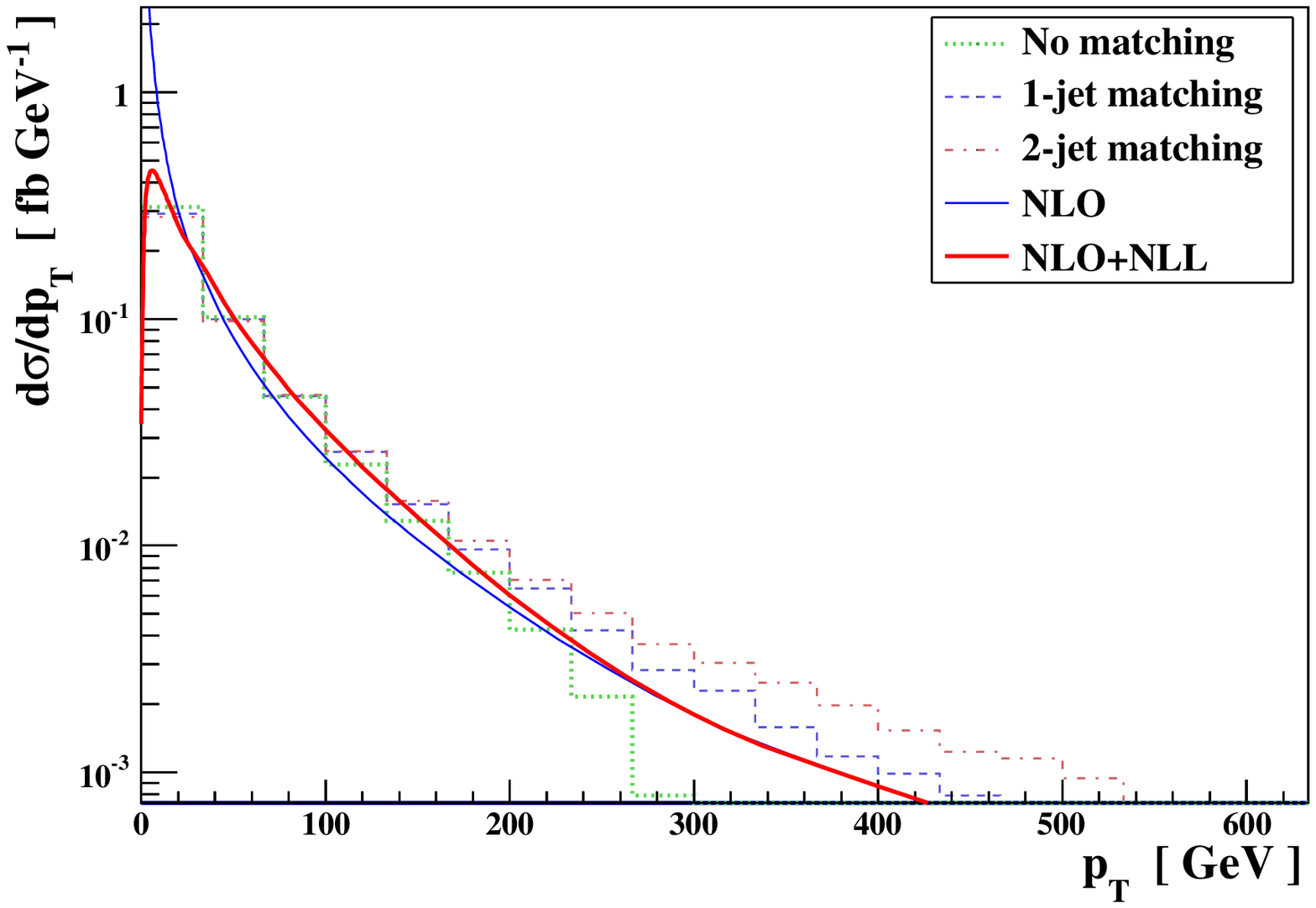,width=.9\textwidth}
 \caption{\label{fig:9}
 Same as Figure \ref{fig:6} for the production of a $\tilde\chi^-_1\tilde\chi^0_2$
associated pair.}
\end{figure}
%

%
\begin{figure}
 \centering
 \epsfig{file=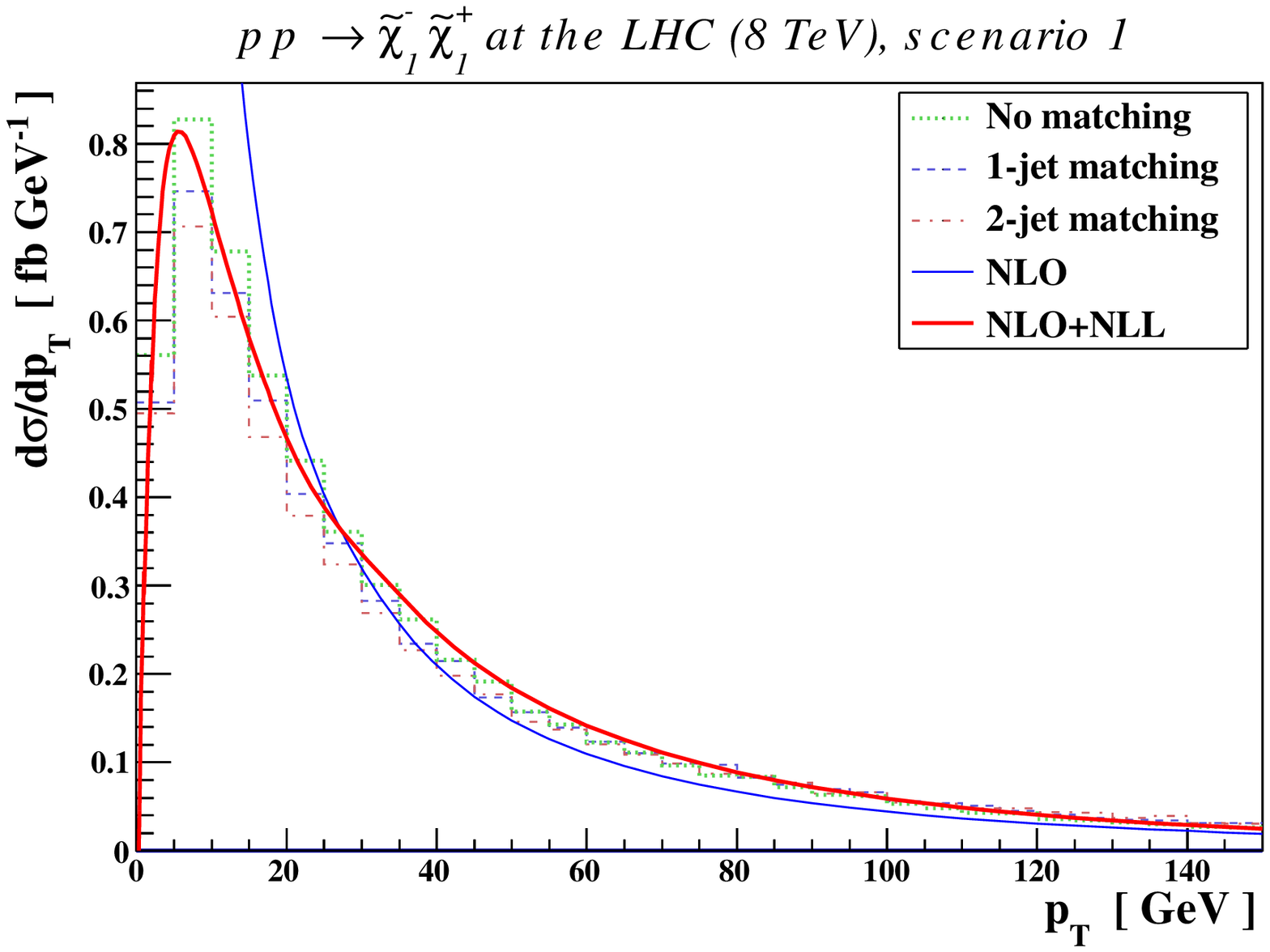,width=.9\textwidth}
 \epsfig{file=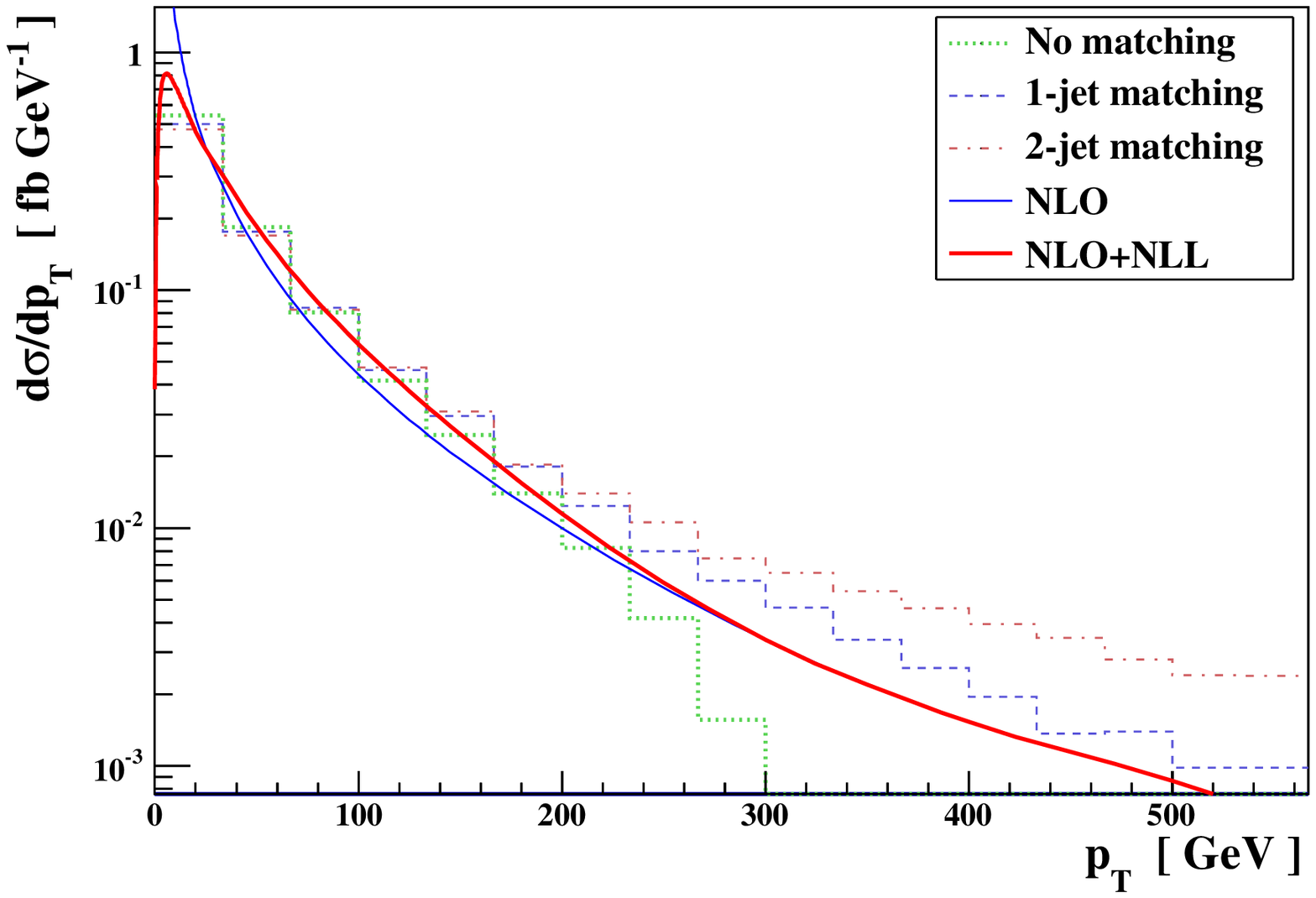,width=.9\textwidth}
 \caption{\label{fig:10}
 Same as Figure \ref{fig:6} for the production of a $\tilde\chi^+_1\tilde\chi^-_1$ pair.}
\end{figure}
%

In Figures \ref{fig:9} and \ref{fig:10}, we address the production of two other
pairs of gauginos, {\it i.e.} of $\tilde\chi_1^-\tilde\chi_2^0$ and $\tilde
\chi_1^+\tilde\chi_1^-$. As
before, the Monte Carlo curves have been normalized according to the resummed
results of Table \ref{tab:4}, {\it i.e.}, to 17.05 fb and 30.04 fb for the
associated production and the chargino pair production channels, respectively.
The distributions are similar to those of the golden channel $\tilde\chi_2^0\tilde\chi_1^+$,
but the absolute size of the latter (see Figure \ref{fig:6}) exceeds the one of the 
$\tilde\chi_1^-\tilde\chi_2^0$ channel (see Figure \ref{fig:9}) as expected for
proton-proton collisions where the production of a positively charged final state 
is favored with respect to a negatively charged final state.

\section{Conclusion}
\label{sec:4}
With a Higgs boson of about 125 GeV to be unveiled, but no hint yet for any
colored supersymmetric particle, experimental searches are now more and more
focusing on the pair production of the superpartners of the gauge and Higgs
bosons.

In this paper, we have updated the total cross sections related to the
production of various gaugino pairs at the LHC, running presently at a center-of-mass
energy of 8 TeV. We have presented results at the leading order and the
next-to-leading order of perturbative QCD and matched, in addition, the NLO
results to threshold resummation at the next-to-leading logarithmic accuracy 
in order to consistently account for the large logarithmic corrections arising
from soft parton emission. Our cross sections have been given for several
benchmark scenarios motivated by the recent LHC supersymmetric searches,
together with the
theoretical uncertainties stemming from scale variation and the choice of the
parton density sets. We have shown that resummation at the NLL accuracy allows
for a drastic reduction of the scale dependence of the cross section to less or
about a percent, in contrast to the PDF errors which are now the dominant source
of theoretical uncertainties and in general of the order of 2--3 percent.

We have also analyzed the precision of the traditional experimental approach
for supersymmetric process simulation consisting in merging parton showering
with multi-parton matrix elements. We have shown that the predictions obtained
following this approach are largely in agreement with the most precise
theoretical computations matching resummation at the NLL level and fixed order
calculations at the NLO accuracy.

We finally emphasize that even if increasing experimental limits on the squark
and gluino masses might exclude the investigated scenarios, the (lower) gaugino
masses may still remain allowed so that our results could stay valid in the
context of more general SUSY-breaking scenarios yielding gaugino masses of the
same order. As already stated above, further numerical results are available
from the authors upon request.

\appendix
\section{Total cross sections at $\sqrt{s}=7$ TeV for benchmark points 1, 18, and 31}
\label{sec:a}

In this Appendix, we present in Tables \ref{tab:7}, \ref{tab:8}, and \ref{tab:9}
the total production cross sections for the same combinations of neutralino and
chargino pairs in the context of the benchmark points 1, 18 and 31 as in Section
\ref{sec:2b}, but now for data taken earlier at the LHC at a center-of-mass
energy of $\sqrt{s}=7$ TeV. We give again numerical results at the leading and
next-to-leading order of perturbative QCD as well as after matching the NLO results
to threshold resummation.

%
\begin{table}[!t]
\renewcommand{\arraystretch}{1.2}
\begin{center}
\begin{tabular}{| l | l | l | l | l | l |}
\hline
 Process & $m_1$ [GeV] & $m_2$ [GeV] &  LO [fb] & NLO [fb] & NLO+NLL [fb] \\
\hline
$p p \to \chi^0_1 \chi^0_1$ & 161.7 & 161.7 & $0.61^{+6.7 \%}_{-6.1 \%}$  &  $0.79^{+3.7 \%}_{-3.2 \%}{}^{+3.0 \%}_{-2.1 \%}$  & $0.77^{+0.4 \%}_{-0.7 \%}{}^{+2.9 \%}_{-2.1 \%}$\\

$p p \to \chi^0_1 \chi^-_1$ & 161.7 & 303.5 & $0.12^{+7.0 \%}_{-6.3 \%}$  &  $0.15^{+2.6 \%}_{-2.6 \%}{}^{+3.1 \%}_{-2.5 \%}$  & $0.14^{+0.0 \%}_{-0.4 \%}{}^{+3.1 \%}_{-2.7 \%}$\\

$p p \to \chi^0_2 \chi^0_2$ & 303.8 & 303.8 & $0.58^{+10.2 \%}_{-8.7 \%}$  &  $0.73^{+3.7 \%}_{-3.8 \%}{}^{+3.5 \%}_{-2.2 \%}$  & $0.71^{+0.1 \%}_{-0.1 \%}{}^{+3.7 \%}_{-2.1 \%}$\\

$p p \to \chi^0_2 \chi^0_3$ & 303.8 & 526.5 & $0.14^{+10.5 \%}_{-9.0 \%}$  &  $0.17^{+2.8 \%}_{-3.2 \%}{}^{+3.6 \%}_{-2.4 \%}$  & $0.17^{+0.2 \%}_{-0.5 \%}{}^{+3.7 \%}_{-2.4 \%}$\\

$p p \to \chi^0_2 \chi^-_1$ & 303.8 & 303.5 & $10.42^{+7.7 \%}_{-6.9 \%}$  &  $12.33^{+1.7 \%}_{-2.0 \%}{}^{+3.3 \%}_{-2.7 \%}$  & $12.18^{+0.3 \%}_{-0.8 \%}{}^{+3.3 \%}_{-2.7 \%}$\\

$p p \to \chi^0_3 \chi^0_4$ & 526.5 & 542.4 & $0.52^{+12.2 \%}_{-10.2 \%}$  &  $0.60^{+3.1 \%}_{-3.6 \%}{}^{+4.4 \%}_{-2.6 \%}$  & $0.59^{+0.6 \%}_{-1.2 \%}{}^{+4.5 \%}_{-2.8 \%}$\\

$p p \to \chi^0_3 \chi^-_2$ & 526.5 & 542.2 & $0.25^{+12.4 \%}_{-10.4 \%}$  &  $0.30^{+3.1 \%}_{-3.6 \%}{}^{+5.7 \%}_{-4.2 \%}$  & $0.30^{+0.7 \%}_{-1.3 \%}{}^{+5.8 \%}_{-4.1 \%}$\\

$p p \to \chi^0_4 \chi^-_2$ & 542.4 & 542.2 & $0.24^{+12.5 \%}_{-10.4 \%}$  &  $0.28^{+2.9 \%}_{-3.5 \%}{}^{+5.8 \%}_{-4.3 \%}$  & $0.28^{+0.8 \%}_{-1.4 \%}{}^{+5.9 \%}_{-4.2 \%}$\\

$p p \to \chi^+_1 \chi^0_1$ & 303.5 & 161.7 & $0.29^{+6.9 \%}_{-6.2 \%}$  &  $0.35^{+2.6 \%}_{-2.6 \%}{}^{+3.0 \%}_{-2.2 \%}$  & $0.35^{+0.1 \%}_{-0.4 \%}{}^{+3.0 \%}_{-2.2 \%}$\\

$p p \to \chi^+_1 \chi^0_2$ & 303.5 & 303.8 & $26.71^{+7.4 \%}_{-6.6 \%}$  &  $30.78^{+1.7 \%}_{-1.9 \%}{}^{+3.2 \%}_{-2.3 \%}$  & $30.47^{+0.0 \%}_{-0.4 \%}{}^{+3.2 \%}_{-2.3 \%}$\\

$p p \to \chi^+_1 \chi^0_3$ & 303.5 & 526.5 & $0.24^{+10.4 \%}_{-8.8 \%}$  &  $0.28^{+2.8 \%}_{-3.2 \%}{}^{+4.0 \%}_{-2.6 \%}$  & $0.27^{+0.0 \%}_{-0.3 \%}{}^{+4.1 \%}_{-2.6 \%}$\\

$p p \to \chi^+_1 \chi^-_1$ & 303.5 & 303.5 & $18.90^{+7.6 \%}_{-6.8 \%}$  &  $22.18^{+1.8 \%}_{-2.1 \%}{}^{+2.9 \%}_{-2.1 \%}$  & $21.93^{+0.1 \%}_{-0.6 \%}{}^{+2.9 \%}_{-2.1 \%}$\\

$p p \to \chi^+_2 \chi^0_3$ & 542.2 & 526.5 & $0.81^{+12.4 \%}_{-10.4 \%}$  &  $0.91^{+3.1 \%}_{-3.7 \%}{}^{+4.9 \%}_{-3.0 \%}$  & $0.91^{+0.5 \%}_{-1.0 \%}{}^{+4.8 \%}_{-3.0 \%}$\\

$p p \to \chi^+_2 \chi^0_4$ & 542.2 & 542.4 & $0.76^{+12.5 \%}_{-10.4 \%}$  &  $0.85^{+3.0 \%}_{-3.6 \%}{}^{+4.9 \%}_{-3.0 \%}$  & $0.85^{+0.6 \%}_{-1.1 \%}{}^{+4.9 \%}_{-3.3 \%}$\\

$p p \to \chi^+_2 \chi^-_2$ & 542.2 & 542.2 & $0.54^{+12.1 \%}_{-10.1 \%}$  &  $0.62^{+2.9 \%}_{-3.5 \%}{}^{+4.6 \%}_{-2.7 \%}$  & $0.61^{+0.6 \%}_{-1.2 \%}{}^{+4.6 \%}_{-2.7 \%}$\\

\hline
\end{tabular}
\caption{\label{tab:7}
 Total cross sections related to the production of various gaugino pairs of masses 
 $m_1$ and $m_2$, presented together with the associated scale and PDF
 uncertainties
 for the LHC running at a center-of-mass energy of $\sqrt{s}=7$ TeV in
 the context of the benchmark point 1 of the LPCC numbering scheme. The cross
 sections are given at the leading order and next-to-leading order of
 perturbative QCD and matched to threshold resummation.
 The PDF uncertainties are not shown for the LO results. Any cross section smaller
 than 0.1 fb is omitted.}
\end{center}
\end{table}
%

Due to the slightly lower available energy, the cross sections are smaller
by about 30\% with respect to those presented in Tables \ref{tab:4}--\ref{tab:6}.
In particular, the cross sections for heavy gaugino pairs do not exceed
0.1 fb at the benchmark points 18 and 31, so that they are no longer listed.
On the other hand, 
the ATLAS and CMS experiments accumulated in 2011 an integrated luminosity
of around 6 fb$^{-1}$ each, so that for the light channels a significant
number of gaugino events could have been produced at all three benchmark
points considered here.

%
\begin{table}[t!]
\renewcommand{\arraystretch}{1.2}
\begin{center}
\begin{tabular}{| l | l | l | l | l | l |}
\hline
 Process & $m_1$ [GeV] & $m_2$ [GeV] &  LO [fb] & NLO [fb] & NLO+NLL [fb]  \\
\hline
$p p \to \chi^0_2 \chi^-_1$ & 471.9 & 471.8 & $1.05^{+11.1 \%}_{-9.4 \%}$  &  $1.21^{+2.0 \%}_{-2.7 \%}{}^{+4.7 \%}_{-3.5 \%}$  & $1.19^{+0.8 \%}_{-1.4 \%}{}^{+4.8 \%}_{-3.4 \%}$\\

$p p \to \chi^+_1 \chi^0_2$ & 471.8 & 471.9 & $3.20^{+11.0 \%}_{-9.3 \%}$  &  $3.53^{+2.0 \%}_{-2.7 \%}{}^{+4.3 \%}_{-2.7 \%}$  & $3.49^{+0.6 \%}_{-1.1 \%}{}^{+4.4 \%}_{-2.8 \%}$\\

$p p \to \chi^+_1 \chi^-_1$ & 471.8 & 471.8 & $2.07^{+11.0 \%}_{-9.3 \%}$  &  $2.33^{+2.1 \%}_{-2.8 \%}{}^{+3.9 \%}_{-2.4 \%}$  & $2.30^{+0.6 \%}_{-1.3 \%}{}^{+3.8 \%}_{-2.5 \%}$\\

\hline
\end{tabular}
\caption{\label{tab:8}
 Same as Table \ref{tab:7} for the benchmark point 18 of the LPCC numbering
scheme.}
\end{center}
\end{table}
%

%
\begin{table}[t!]
\renewcommand{\arraystretch}{1.2}
\begin{center}
\begin{tabular}{| l | l | l | l | l | l |}
\hline
 Process & $m_1$ [GeV] & $m_2$ [GeV] &  LO [fb] & NLO [fb] & NLO+NLL [fb]  \\
\hline
$p p \to \chi^0_2 \chi^-_1$ & 478.5 & 478.5 & $0.96^{+11.2 \%}_{-9.5 \%}$  &  $1.10^{+2.0 \%}_{-2.7 \%}{}^{+4.8 \%}_{-3.5 \%}$  & $1.08^{+0.9 \%}_{-1.5 \%}{}^{+4.6 \%}_{-3.6 \%}$\\

$p p \to \chi^+_1 \chi^0_2$ & 478.5 & 478.5 & $2.94^{+11.1 \%}_{-9.4 \%}$  &  $3.23^{+2.0 \%}_{-2.7 \%}{}^{+4.4 \%}_{-2.7 \%}$  & $3.20^{+0.4 \%}_{-1.1 \%}{}^{+4.6 \%}_{-2.8 \%}$\\

$p p \to \chi^+_1 \chi^-_1$ & 478.5 & 478.5 & $1.90^{+11.1 \%}_{-9.4 \%}$  &  $2.13^{+2.1 \%}_{-2.8 \%}{}^{+3.9 \%}_{-2.4 \%}$  & $2.11^{+0.5 \%}_{-1.1 \%}{}^{+3.8 \%}_{-2.6 \%}$\\

\hline
\end{tabular}
\caption{\label{tab:9}
 Same as Table \ref{tab:7} for the benchmark point 31 of the LPCC numbering
scheme.}
\end{center}
\end{table}
%

\acknowledgments

The authors are grateful to Johan Alwall and Olivier Mattelaer for their help
with the {\sc MadGraph} program. This work has been supported by the BMBF
Theorie-Verbund and by the Theory-LHC-France initiative of the CNRS/IN2P3. 


\end{document}